\documentclass[a4paper,11pt]{article}
\usepackage{jcappub}
\usepackage{lineno}

\usepackage{graphicx}
\usepackage{amsmath,amssymb}
\usepackage{verbatim}
\usepackage{color}
\usepackage{booktabs}
\usepackage{hyperref}
\usepackage{fancyhdr, fancybox, empheq}
\usepackage{subfigure}
\usepackage{float}
\usepackage{url}
\usepackage[dvipsnames]{xcolor}
\usepackage{mathtools}
\usepackage[normalem]{ulem}
\usepackage{comment}
\usepackage{xspace}
\usepackage{bm}
\usepackage{multirow}

\hypersetup{
 colorlinks=true,
 urlcolor=blue,
 linkcolor=blue,
 citecolor=blue
}

\newcommand{\agora}{\textsc{Agora}}
\newcommand{\camb}{\textsc{CAMB}}

\newcommand{\planck}{{\it Planck}}

\newcommand{\polarbear}{\textsc{Polarbear}}
\newcommand{\keckarray}{\textit{Keck Array}}
\newcommand{\mh}{tSZ-deproj}

\arxivnumber{}
\title{Foreground Mitigation for CMB Lensing with the Global Minimum Variance Quadratic Estimator}

\author[a,1]{Yuka Nakato\note{Corresponding author.},}
\author[b,c]{W.L. Kimmy Wu,}
\author[a,c]{Ana Carolina Silva Oliveira,}
\author[d,e,f]{Yuuki Omori,}
\author[a,b]{and Abhishek S. Maniyar}
\affiliation[a]{Department of Physics, Stanford University,\\
382 Via Pueblo Mall, Stanford, CA 94305, USA}
\affiliation[b]{SLAC National Accelerator Laboratory,\\2575 Sand Hill Road, Menlo Park, CA 94025, USA}
\affiliation[c]{California Institute of Technology,\\
1200 E. California Boulevard, Pasadena, CA 91125, USA}
\affiliation[d]{Kavli Institute for Cosmological Physics, University of Chicago,\\
5640 South Ellis Avenue, Chicago, IL 60637, USA}
\affiliation[e]{Department of Astronomy and Astrophysics, University of Chicago,\\
5640 South Ellis Avenue, Chicago, IL, 60637, USA}
\affiliation[f]{NSF-Simons AI Institute for the Sky (SkAI),\\
172 E. Chestnut St., Chicago, IL 60611, USA}
\emailAdd{yukanaka@stanford.edu}

\abstract{
Weak gravitational lensing of the cosmic microwave background (CMB) is a powerful probe of cosmology, providing insight into structure formation and the evolution of the universe. 
Current and upcoming CMB experiments such as SPT-3G and the Simons Observatory (SO) provide high-resolution, low-noise temperature and polarization maps that are ideal for lensing reconstruction. 
The global minimum variance (GMV) quadratic estimator for CMB lensing reduces reconstruction noise  over the standard quadratic estimator (SQE).
In this work, we extend the GMV framework to incorporate the \mh\ and cross-ILC foreground mitigation techniques, which enhance robustness against contamination from astrophysical sources.
For a simulation study using SPT-3G Ext-10k and SO Extended configurations at $\ell_{\mathrm{max}}^T = 3500$, the lensing bias at $L < 1000$ is reduced from $\sim4\%$ with standard GMV and SQE to $2\%$ with \mh, and to $< 1\%$ with cross-ILC.
These methods enable the construction of foreground-cleaned lensing maps suitable for cross-correlation analyses, with direct relevance for current and future surveys.
}

\begin{document}
\maketitle
\flushbottom

\section{Introduction} \label{sec:introduction}

Gravitational lensing perturbs the trajectories of cosmic microwave background (CMB) photons as they propagate from the surface of last scattering, inducing characteristic correlations between modes in the observed temperature and polarization maps.
Recent measurements of CMB lensing have been reported with \textit{Planck} data \cite{planck_pr4_lensing}, Atacama Cosmology Telescope (ACT)~\cite{act_dr6_lensing}, the South Pole Telescope using its third-generation receiver (SPT-3G)~\cite{spt3g_muse_lensing, spt_lensing_20192020}, \polarbear~\cite{polarbear_lensing_2020}, and BICEP2/\keckarray~\cite{bicep_lensing_2016}.
Looking ahead, experiments like the Simons Observatory~(SO) will deliver lower noise levels and wider sky coverage, improving the precision of lensing measurements. These gains make it increasingly important to develop improved methods for lensing reconstruction, as CMB lensing will play a central role in constraining fundamental physics, including neutrino masses and the growth of structure \cite{lewis_challinor_2006}.

The quadratic estimator (QE) \cite{hu_okamoto_2002, okamoto_hu_2003} reconstructs the lensing potential $\hat{\phi}$ from lensing-induced mode couplings in the CMB, using two input CMB maps. We write this as $\hat{\phi} \equiv \hat{\phi}(X,Y)$, where the fields $X$ and $Y$ are drawn from the CMB temperature ($T$) and polarization ($E,B$) maps.
From this reconstruction, we can construct the lensing angular power spectrum $C_L^{\phi\phi}$, one of the key observables used to constrain cosmological parameters.
The first version of this method was introduced by Hu \& Okamoto in \cite{hu_okamoto_2002}, which we call the standard quadratic estimator (SQE). 
In the SQE framework, the lensing estimator is constructed individually for each map pair $TT,EE,TE,EB$, and $TB$.
The minimum-variance (MV) estimator is then obtained by forming a linear combination of these five pairings, where the weights minimize the total variance. However, this method is not quite optimal because it performs the filtering and construction of each $XY$ estimator independently before combining them, rather than optimizing all estimators jointly within a unified framework.

The global minimum variance (GMV) quadratic estimator, introduced in \cite{hirata_and_seljak, gmv}, improves upon the SQE and has been implemented under the curved-sky formalism in the \planck\ PR4 lensing analysis~\cite{planck_pr4_lensing} and the SPT-3G D1 lensing analysis, the most recent SPT-3G lensing analysis using data from the Main 1500 deg$^2$ field observed during the 2019 and 2020 austral winter seasons~\cite{spt_lensing_20192020}.
Unlike the SQE, which constructs and filters each $XY$ estimator independently, the GMV estimator performs a joint inverse-variance filtering of the CMB temperature and polarization input maps, accounting for all covariances between $T$, $E$, and $B$.
In sensitive experiments such as SO and SPT-3G, the GMV estimator yields up to about a 10\% improvement in reconstruction noise compared to SQE implementations \cite{gmv}.

One of the biggest challenges for CMB lensing reconstruction is contamination from astrophysical foregrounds. Extragalactic signals such as the thermal Sunyaev-Zel'dovich (tSZ) effect, kinetic Sunyaev-Zel'dovich (kSZ) effect, cosmic infrared background (CIB), and radio sources can contribute to the observed CMB signal.
The tSZ effect arises when low-energy CMB photons collide with hot electrons that mostly reside in galaxy clusters and receive an energy boost via inverse Compton scattering, resulting in a characteristic distortion of the CMB spectrum \cite{sz_1970, sz_1972}. 
tSZ is responsible for the majority of the foreground bias in temperature data. 
The kSZ originates from the scattering of photons off free electrons in the intergalactic medium moving with non-zero bulk velocity relative to the CMB rest frame~\cite{sz_1980}.
The CIB is another foreground, originating from integrated infrared emission of dusty star-forming galaxies \cite{puget_1996, fixsen_1998}. Radio sources such as active galactic nuclei (AGN) and radio galaxies can also contaminate the CMB signal \cite{toffolatti_1998}.
Among the remaining foregrounds after typical masking or subtraction of bright point sources and large clusters in CMB lensing analyses, the tSZ and CIB are the dominant sources of bias in the lensing measurement.

All of these extragalactic foregrounds have non-Gaussian probability densities and, because many of them trace the same large-scale structure that lenses the CMB, they are correlated with the true lensing potential. This leads to bispectrum-type contributions (three-point correlations), such as those involving $\phi$ and two foreground fields. In addition, the intrinsic non-Gaussianity of the foregrounds themselves produces a foreground trispectrum (connected four-point function).
These contributions can mimic lensing-induced mode couplings and bias the reconstructed lensing power spectrum. The impact of these foregrounds is especially significant in the $TT$ lensing estimator, as they are largely unpolarized \cite{gupta_2019, datta_2019} and thus contribute primarily to temperature maps.
Current CMB lensing analyses such as \planck\ and ACT rely heavily on the $TT$ estimator and achieve two-percent-level constraints on the lensing amplitude \cite{planck_pr4_lensing, act_dr6_lensing}.
At this level of statistical sensitivity, biases from extragalactic foregrounds become a critical concern: their contributions to the four-point function can reach the several-percent level in temperature-based reconstructions if unmitigated, and they are difficult to model accurately. Thus, to ensure that foreground biases remain subdominant to statistical errors, they must be reduced to below the percent level, motivating the development of dedicated mitigation strategies \cite{mh_2018, crossilc_2023, namikawa_hanson_takahashi_2013, osborne_hanson_dore_2014, sailer_schaan_ferraro_2020}.

In this paper, we incorporate two foreground mitigation methods, \mh\ \cite{mh_2018} and cross-ILC \cite{crossilc_2023}, into the GMV framework to enhance the robustness of CMB lensing reconstruction against contamination from extragalactic sources. These techniques were previously introduced in the SQE formalism using only the $TT$ estimator. Here, we show that they can be generalized to the full GMV formalism, and demonstrate their performance for current and upcoming CMB experiments. These techniques were also applied in the SPT-3G D1 lensing analysis, together with profile hardening techniques \cite{namikawa_hanson_takahashi_2013, osborne_hanson_dore_2014, sailer_schaan_ferraro_2020}, which we do not discuss in detail here.

\mh\ and cross-ILC both mitigate foreground-induced biases by reducing the overlap of the same non-Gaussian foreground residuals between the two inputs of the quadratic estimator.
This suppression at the map level prevents spurious bispectrum and trispectrum contributions from propagating into the four-point function used in lensing reconstruction. Reducing this bias allows us to use higher maximum multipoles ($\ell_{\mathrm{max}}$) from the temperature maps while keeping the $TT$ estimator unbiased,
improving the overall signal-to-noise ratio of the lensing spectrum measurement, especially in the mildly non-linear regime \cite{hu_okamoto_2002}.

This paper is organized as follows. We begin in Section~\ref{sec:mh_and_crossilc_formalism} by presenting the derivation and implementation of the \mh\ and cross-ILC foreground-immune equations in both the SQE and GMV formalisms.
We then apply these methods to simulations to quantify the resulting bias in the reconstructed lensing spectrum for different experimental configurations.
In Section~\ref{sec:inputs} we describe the simulations used as inputs in this work. In Section~\ref{sec:numerical_results} we present the numerical results and discuss the performance of each mitigation technique. We then conclude in Section~\ref{sec:conclusion}.

\section{tSZ-Deproj and Cross-ILC Formalism} \label{sec:mh_and_crossilc_formalism}

The lensing-induced correlations between modes in the observed CMB maps can be written as
\begin{equation} \label{eq:lensing}
	\langle X_{\ell m} Y_{\ell' m'} \rangle_{\mathrm{CMB}} = \sum_{LM} (-1)^M
	\begin{pmatrix}
		\ell & \ell' & L\\
		m & m' & -M\\
	\end{pmatrix}
	f^{XY}_{\ell \ell' L} \phi_{LM},
\end{equation}
where $X,Y \in \{T,E,B\}$ are the lensed temperature and polarization fields, $\phi$ is the lensing potential, $f^{XY}_{\ell \ell' L}$ is a weight representing the lensing response to the CMB which is fixed for each choice of $XY$ pairing (functional forms of $f^{XY}_{\ell \ell' L}$ can be found in \cite{okamoto_hu_2003}), and $\langle \cdot \rangle_{\rm CMB}$ denotes average over CMB realizations at fixed $\phi$.
Here we assume $L > 0$ \cite{okamoto_hu_2003}.

Both the \mh\ and cross-ILC methods rely on asymmetric lensing reconstruction, in which the two input maps $X$ and $Y$ are intentionally chosen to differ in type. For instance, for the $TT$ estimator, typically the two inputs of the estimator use the same temperature map. However, in the \mh\ method introduced by \cite{mh_2018}, one of the temperature inputs is replaced with a tSZ-nulled internal linear combination (ILC) map, while the other remains the standard minimum-variance ILC (MV ILC) temperature map.
This configuration nulls tSZ-induced bias in the resulting estimator, with only a small penalty in reconstruction noise.
The tSZ-nulled map does contain a modest level of enhanced residual CIB due to the spectral dissimilarity between the two components, but this effect contributes only weakly to the overall lensing bias compared to the much larger tSZ-induced bias that is eliminated.
Cross-ILC \cite{crossilc_2023} adopts a similar approach, but the two inputs to the quadratic estimator are chosen to be a tSZ-nulled ILC map and a CIB-nulled ILC map. This suppresses biases from both tSZ and CIB, but comes at the cost of increased reconstruction noise, because the CIB-nulled map has higher variance than the MV ILC map, making the penalty larger than in the \mh\ case.

In the following subsections, we first introduce \mh\ and cross-ILC in the simple SQE $TT$-only case, and show how the foreground biases are reduced. Next, we review the standard GMV formalism with no foreground treatment, and its key differences from SQE. Finally, we derive the \mh\ and cross-ILC equations in the GMV formalism.

\subsection{tSZ-Deproj and Cross-ILC SQE \texorpdfstring{$TT$}{TT} Derivation}

In the curved-sky SQE formalism~\cite{okamoto_hu_2003}, the $TT$ quadratic estimator takes the form
\begin{align}\label{eq:sqe_phi_TT}
    \hat{\phi}^{TT}_{LM} &= \frac{\lambda^{TT}(L)}{2} \sum_{\ell m} \sum_{\ell' m'} (-1)^M
	\begin{pmatrix}
		\ell & \ell' & L\\
		m & m' & -M\\
	\end{pmatrix}
	f^{TT}_{\ell \ell' L} \overline{T}_{\ell m} \overline{T}_{\ell' m'}.
\end{align}
Here, $\overline{T}$ denotes the inverse-variance filtered $T$ field
\begin{equation}\label{eq:sqe_tbar}
	\overline{T}_{\ell m} = \frac{T_{\ell m}}{C_\ell^{TT}}  
\end{equation}
where $C_\ell^{TT}$ is the total observed temperature power spectrum (signal plus noise), and $f^{TT}_{\ell \ell' L}$ is the CMB lensing correlation coefficient which is the same one that appears in Eq.~\ref{eq:lensing}. The Lagrange multiplier $\lambda^{TT}$~\cite{gmv} normalizes the estimator such that it is unbiased, i.e., $\langle \hat{\phi}^{TT}_{LM} \rangle_{\rm{CMB}} = \phi_{LM}$. Explicitly, $\lambda^{TT}(L)$ is the inverse of the response function:
\begin{equation}\label{eq:sqe_lambda_TT}
    \lambda^{TT}(L) = \left[ \frac{1}{2(2L+1)} \sum_{\ell \ell'} \frac{|f^{TT}_{\ell \ell' L}|^2}{C_\ell^{TT} C_{\ell'}^{TT}} \right]^{-1}.
\end{equation}

In the \mh\ and cross-ILC estimators, two different $T$ maps are used as inputs. In the \mh\ case, $T_1$ is the tSZ-nulled temperature map, and  $T_2$ is the MV ILC temperature map.
In cross-ILC, $T_1$ is again the tSZ-nulled map, and $T_2$ is the the CIB-nulled map.
The use of two distinct temperature inputs helps suppress biases from correlated foreground residuals, since foreground contamination common to both maps does not contribute to the cross-estimator. To ensure that the estimator remains unbiased under exchange of the two inputs, we symmetrize it as
\begin{equation}\label{eq:sqe_T1T2_plus_T2T1}
    \hat{\phi}^{TT}_{LM} = \frac{1}{2} \left(\hat{\phi}^{T_1T_2}_{LM} + \hat{\phi}^{T_2T_1}_{LM}\right),
\end{equation}
where $\hat{\phi}^{T_1T_2}_{LM}$ denotes the reconstruction using $T_1$ for the first temperature field and $T_2$ for the second, and vice versa for $\hat{\phi}^{T_2T_1}_{LM}$.

To understand how this asymmetric reconstruction technique suppresses foreground-induced biases, we first review how the lensing power spectrum is constructed. The quadratic estimator extracts lensing from mode couplings, i.e., two-point correlations. Then the observable lensing power spectrum $C_L^{\phi\phi}$ comes from correlating lensing maps reconstructed using two quadratic estimators, i.e., it is obtained from the four-point function of the observed CMB fields. This four-point function contains not only the desired true lensing trispectrum, but also intrinsic $N_{L}^{(0)}$ and $N_{L}^{(1)}$ biases. The $N_{L}^{(0)}$ bias is the Gaussian disconnected part of the four-point function, which arises even in the absense of lensing. $N_{L}^{(1)}$ is the subdominant connected contribution, arising from lensing contractions beyond the leading order. These intrinsic biases are estimated and subtracted out using simulations. In addition, we have foreground non-Gaussianities that mimic lensing, which can bias the reconstructed $C_L^{\phi\phi}$ if not suppressed.

We now express the observed temperature maps as the sum of a lensed CMB component and a foreground component, and explicitly expand the four-point function of these maps to examine how \mh\ and cross-ILC suppress foreground-induced terms. We write the connected part of the four-point function schematically as
\begin{align}\label{eq:TTTT}
    \langle TTTT \rangle_c &= \langle (\tilde{T}+T^{\mathrm{fg,A}}) (\tilde{T}+T^{\mathrm{fg,B}}) (\tilde{T}+T^{\mathrm{fg,A}}) (\tilde{T}+T^{\mathrm{fg,B}}) \rangle_c \nonumber\\
    &= \langle \tilde{T}\tilde{T}\tilde{T}\tilde{T} \rangle_c \ \textnormal{(``primary CMB trispectrum")} \nonumber\\
    &\hspace{0.3cm} + \langle T^{\mathrm{fg,A}}T^{\mathrm{fg,B}}T^{\mathrm{fg,A}}T^{\mathrm{fg,B}} \rangle_c \ \textnormal{(``foreground  trispectrum")} \nonumber\\
    &\hspace{0.3cm} + \langle \tilde{T}\tilde{T}\tilde{T}T^{\mathrm{fg,B}} \rangle_c + \cdots \nonumber\\
    &\hspace{0.3cm} + \langle \tilde{T}\tilde{T}T^{\mathrm{fg,A}}T^{\mathrm{fg,B}} \rangle_c + \cdots \nonumber\\
    &\hspace{0.3cm} + \langle \tilde{T}T^{\mathrm{fg,B}}T^{\mathrm{fg,A}}T^{\mathrm{fg,B}} \rangle_c + \cdots
\end{align}
where $\tilde{T}$ denotes the lensed CMB temperature map, and $T^{\mathrm{fg,A}}$ and $T^{\mathrm{fg,B}}$ represent foreground components of the two input legs of the estimator. We focus only on the connected part of the four-point function $\langle TTTT \rangle$, since the disconnected contractions do not bias the lensing reconstruction; these are captured in the $N_{L}^{(0)}$ bias estimated from Gaussian simulations and subtracted out.
The full expansion of Eq.~\ref{eq:TTTT} including all permutations gives 16 total connected four-point terms, which we describe below.

For this discussion, we assume that the unlensed CMB and the foreground fields are independent, and similarly for the unlensed CMB and the lensing potential (i.e., we ignore the small correlations between the integrated Sachs-Wolfe (ISW) signal and large-scale structure fields).
Under this assumption, all terms containing three factors of $\tilde{T}$ or $T^{\mathrm{fg}}$ (i.e., terms like $\langle \tilde{T}\tilde{T}\tilde{T}T^{\mathrm{fg,B}} \rangle_c$ or $ \langle \tilde{T}T^{\mathrm{fg,B}}T^{\mathrm{fg,A}}T^{\mathrm{fg,B}} \rangle_c$ and permutations) vanish.

The first term, $\langle \tilde{T}\tilde{T}\tilde{T}\tilde{T} \rangle_c$, is the primary CMB trispectrum, which contains the true lensing signal and yields the lensing power spectrum $C_L^{\phi\phi}$. This term is unaffected by the use of asymmetric lensing reconstruction.

Next, we have the foreground trispectrum $\langle T^{\mathrm{fg,A}}T^{\mathrm{fg,B}}T^{\mathrm{fg,A}}T^{\mathrm{fg,B}} \rangle_c$, which biases the lensing reconstruction. This term is suppressed by the \mh\ and cross-ILC methods, as they effectively remove tSZ and CIB, which are the dominant foregrounds.  However, it does not vanish entirely, since residual foregrounds such as kSZ and radio sources are not explicitly nulled by component separation targeting tSZ and CIB. Additionally, CIB-nulled ILC maps cannot remove CIB perfectly because its SED is imperfectly known.

We also encounter six terms with two factors each of $\tilde{T}$ and $T^{\mathrm{fg}}$. 
The two terms in which both $\tilde{T}$ fields enter the same estimator, $\langle \tilde{T}\tilde{T}T^{\mathrm{fg,A}}T^{\mathrm{fg,B}} \rangle_c$ and  $\langle T^{\mathrm{fg,A}}T^{\mathrm{fg,B}}\tilde{T}\tilde{T} \rangle_c$, are called the ``primary bispectrum" contribution.
The remaining four terms where each estimator contains one $\tilde{T}$ and one $T^{\mathrm{fg}}$, e.g. $\langle T^{\mathrm{fg,A}}\tilde{T}\tilde{T}T^{\mathrm{fg,B}} \rangle_c$ and permutations, correspond to the ``secondary bispectrum". 
Both of these terms are referred to as bispectrum biases because the pair of $\tilde{T}$ fields either forms the lensing potential directly (primary bispectrum) or a quantity proportional to the lensing potential (secondary bispectrum), making the terms proportional to the bispectrum $B^{\phi T^{\mathrm{fg}} T^{\mathrm{fg}}}$.
Both bispectrum biases contribute to the lensing bias but are mitigated by the use of \mh\ or cross-ILC. These methods reduce the overlap of foreground content between $T^{\mathrm{fg,A}}$ and $T^{\mathrm{fg,B}}$, suppressing $\langle T^{\mathrm{fg,A}}T^{\mathrm{fg,B}} \rangle$, and also reduce the amplitude of foreground residuals in each map individually, suppressing $\langle \tilde{T}T^{\mathrm{fg}} \rangle$.

To summarize, the leading sources of lensing bias arise from the foreground trispectrum, the primary bispectrum, and the secondary bispectrum. Each of these contributions is suppressed by asymmetric reconstruction methods such as \mh\ and cross-ILC, which reduce correlations between foreground components in the two inputs of the estimator. As a result, asymmetric reconstruction yields a more robust estimate of the lensing power spectrum in the presence of foreground contamination.

\begin{figure}[t]
	\centering
	\includegraphics[width=.6\textwidth]{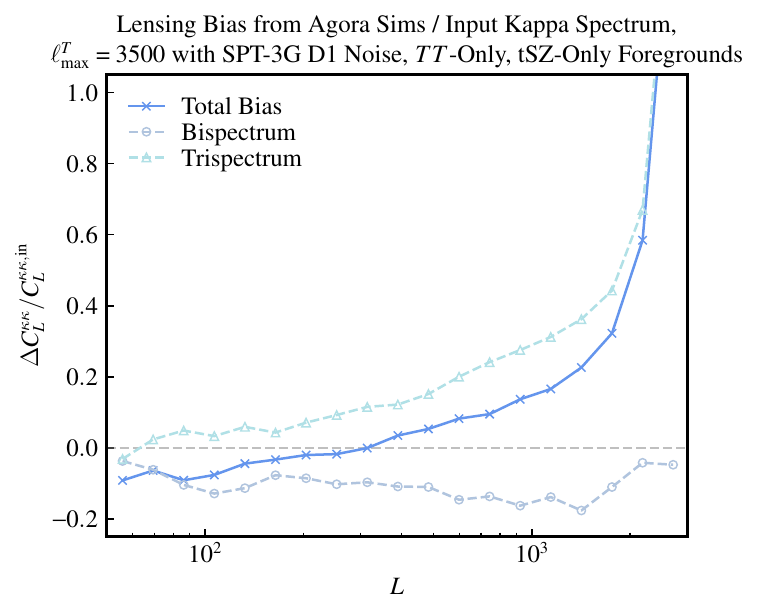}
	\centering \caption{The lensing bias for the $TT$-only SQE reconstruction with no foreground treatment, on input maps containing only CMB and tSZ, assuming experimental noise levels matching the SPT-3G D1 analysis. Here, the lensing bias is defined as $\Delta C_L^{\kappa\kappa} = C_L^{\kappa\kappa,\mathrm{recon}} - C_L^{\kappa\kappa,\mathrm{in}}$ and it is normalized by the input $\kappa$ spectrum. The bias is split up into the trispectrum contribution and the bispectrum contribution.
	The total bias is negative for $L \lesssim 300$, after which it becomes positive as the trispectrum contribution starts to dominate.}
	\label{fig:bias_split_up_agora_standard_sqe}
\end{figure}

Figure~\ref{fig:bias_split_up_agora_standard_sqe} shows the lensing bias for the standard (no foreground treatment) SQE $TT$ reconstruction, using input maps containing lensed CMB and tSZ. To separate the contributions from the foreground trispectrum and the $\kappa$-tSZ-tSZ bispectrum, we decompose the total bias following the method outlined in Section~IV~E of \cite{spt_lensing_2018}. The total bias is defined as
\begin{equation}\label{eq:total_bias}
    \Delta C_L^{\kappa\kappa} = C_L^{\kappa\kappa}(\mathrm{CMB}^\mathrm{NG}+\mathrm{FG}^\mathrm{NG}) - C_L^{\kappa\kappa,\mathrm{in}},
\end{equation}
where $C_L^{\kappa\kappa,\mathrm{in}}$ is the input lensing spectrum, and $C_L^{\kappa\kappa}(\mathrm{CMB}^\mathrm{NG}+\mathrm{FG}^\mathrm{NG})$ is the reconstructed lensing spectrum after applying the response correction and subtracting out the $N_{L}^{(0)}$ and $N_{L}^{(1)}$ biases. Here, $\mathrm{CMB}^\mathrm{NG}$ denotes the lensed CMB simulation, and $\mathrm{FG}^\mathrm{NG}$ is the non-Gaussian foreground simulation from \agora.
The trispectrum component is estimated as
\begin{equation}\label{eq:trispec}
    C_L^{\kappa\kappa}(\mathrm{CMB}^{\mathrm{NG},2}+\mathrm{FG}^\mathrm{NG}) - C_L^{\kappa\kappa,\mathrm{in}},
\end{equation}
where $\mathrm{CMB}^{\mathrm{NG},2}$ is the lensed CMB map rotated relative to the foreground map, so that its lensing field is uncorrelated with the foregrounds.
The bispectrum contribution is then obtained as the difference between the total bias and the trispectrum bias:
\begin{equation}\label{eq:bispec}
    C_L^{\kappa\kappa}(\mathrm{CMB}^\mathrm{NG}+\mathrm{FG}^\mathrm{NG}) - C_L^{\kappa\kappa}(\mathrm{CMB}^{\mathrm{NG},2}+\mathrm{FG}^\mathrm{NG}).
\end{equation}
We find that the total bias is negative at lower $L$s, then changes sign around $L \sim 300$. The trispectrum component is positive as expected (with a small dip in the lowest $L$ bin due to noise scatter).

\subsection{Standard GMV Formalism}
\label{sec:standard_gmv_formalism}

We briefly review the standard GMV formalism as introduced in \cite{gmv}.
Let $\bm{X}_{\ell m} = [T_{\ell m}, E_{\ell m}, B_{\ell m}]$ denote the three-dimensional vector of observed CMB temperature and polarization fields.
Define
\begin{equation}\label{eq:d_ell}
	D_\ell \equiv C_\ell^{TT}C_\ell^{EE} - \left[C_\ell^{TE}\right]^2,
\end{equation}
and assume $C_\ell^{TB} = C_\ell^{BT} = C_\ell^{EB} = C_\ell^{BE}=0$, so that the inverse of the covariance matrix of $\bm{X}_{\ell m}$ is
\begin{align}\label{eq:invcl}
	\bm{C}_\ell^{-1}
	=
	\begin{bmatrix}
		C_\ell^{TT} & C_\ell^{TE} & 0\\
		C_\ell^{ET} & C_\ell^{EE} & 0\\
		0 & 0 & C_\ell^{BB}\\
	\end{bmatrix}^{-1}
	=
	\begin{bmatrix}
		C_\ell^{EE}/D_\ell & -C_\ell^{TE}/D_\ell & 0\\
		-C_\ell^{ET}/D_\ell &C_\ell^{TT}/D_\ell & 0\\
		0 & 0 & 1/C_\ell^{BB}\\
	\end{bmatrix}.
\end{align}

We now define the inverse-variance weighted fields as
\begin{equation}\label{eq:xbar}
	\overline{\bm{X}}_{\ell m} \equiv \bm{C}_\ell^{-1}\bm{X}_{\ell m}.
\end{equation}
Using Eq.~\ref{eq:invcl}, this becomes
\begin{align}\label{eq:xbar_expanded}
	\begin{bmatrix}
		\overline{T}_{\ell m}\\
		\overline{E}_{\ell m}\\
		\overline{B}_{\ell m}\\
	\end{bmatrix}
	=
	\begin{bmatrix}
		C_\ell^{EE}/D_\ell & -C_\ell^{TE}/D_\ell & 0\\
		-C_\ell^{ET}/D_\ell & C_\ell^{TT}/D_\ell & 0\\
		0 & 0 & 1/C_\ell^{BB}\\
	\end{bmatrix}
	\begin{bmatrix}
		T_{\ell m}\\
		E_{\ell m}\\
		B_{\ell m}\\
	\end{bmatrix}
	=
	\begin{bmatrix}
		\left(C_\ell^{EE}T_{\ell m} - C_\ell^{TE}E_{\ell m}\right)/D_\ell\\
		\left(C_\ell^{TT}E_{\ell m} - C_\ell^{ET}T_{\ell m}\right)/D_\ell\\
		B_{\ell m}/C_\ell^{BB}\\
	\end{bmatrix}.
\end{align}
The GMV lensing estimator is then given by
\begin{align}\label{eq:eq52}
	\hat{\phi}_{LM}^{GMV} &= \frac{\lambda(L)}{2} \sum_{\ell m} \sum_{\ell' m'} (-1)^M
    \begin{pmatrix}
		\ell & \ell' & L\\
		m & m' & -M\\
	\end{pmatrix}
    \overline{\bm{X}}^{T}_{\ell m} \bm{f}_{\ell \ell' L} \overline{\bm{X}}_{\ell' m'}.
\end{align}
Here, $\bm{f}_{\ell \ell' L}$ is the full matrix containing all correlation coefficients:
\begin{equation}
    \bm{f}_{\ell \ell' L} =
    \begin{bmatrix}
		f^{TT} & f^{TE} & f^{TB}\\
		f^{TE} & f^{EE} & f^{EB}\\
		f^{TB} & f^{EB} & f^{BB}\\
	\end{bmatrix}.
\end{equation}

Now, there are a few key differences to note between the SQE and GMV formalisms. First, in the SQE $TT$ quadratic estimator shown in Eq.~\ref{eq:sqe_phi_TT}, the inverse-variance filtered temperature field is obtained by simply dividing each $T_{\ell m}$ mode by its corresponding power spectrum $C_{\ell}^{TT}$ (Eq.~\ref{eq:sqe_tbar}). In contrast, the GMV formalism jointly filters the temperature and polarization fields using the full covariance matrix, as seen in Eq.~\ref{eq:xbar_expanded}. The resulting filtered temperature field $\overline{T}$ in the GMV case includes contributions from both $T$ and $E$, reflecting the fact that the filtering is performed collectively rather than field-by-field.

Second, the MV estimator is constructed differently. In SQE, single-pair estimators $\hat{\phi}^{XY}$ are computed individually, and then linearly combined:
\begin{equation} \label{eq:sqe_mv}
	\hat{\phi}^{SQE} = \sum_{XY} w^{XY} \hat{\phi}^{XY},
\end{equation}
where the weights $w^{XY}$ are chosen to minimize the total variance under the constraint $\sum_{XY} w^{XY} = 1$.
In contrast, the GMV estimator performs this optimization jointly, constructing a single estimator that simultaneously incorporates all $T$, $E$, and $B$ information, as shown in Eq.~\ref{eq:eq52}.

This also means that the normalization factor $\lambda$ differs. SQE has a different $\lambda^{XY}$ for each estimator pair, which depends only on the lensing correlation coefficient $f^{XY}$ and the total observed power spectrum $C_{\ell}^{XY}$ for that pair (e.g., Eq.~\ref{eq:sqe_lambda_TT}). On the other hand, GMV has a single $\lambda$ that normalizes the joint estimator, derived from the full covariance matrix $\bm{C}_\ell$ and the matrix of all correlation coefficients $\bm{f}_{\ell \ell' L}$:
\begin{equation}\label{eq:eq43}
    \lambda(L) = \left[ \frac{1}{2(2L+1)} \sum_{\ell \ell'} \mathrm{Tr} \left(\bm{C}_{\ell}^{-1}\bm{f}_{\ell \ell' L}\bm{C}_{\ell'}^{-1}\bm{f}_{\ell' \ell L}\right) \right]^{-1}.
\end{equation}

\subsection{tSZ-Deproj and Cross-ILC GMV Derivation}

To understand how \mh\ and cross-ILC extend to the GMV estimator, we start from the standard GMV expression in Eq.~\ref{eq:eq52}, where the inverse-variance filtered vector is given by $\overline{\bm{X}}_{\ell m} = [\overline{T}_{\ell m}, \overline{E}_{\ell m}, \overline{B}_{\ell m}]$. In the asymmetric case of \mh\ and cross-ILC, however, we now work with two distinct temperature maps $T_1$ and $T_2$, which are separately foreground-cleaned (e.g., tSZ-nulled and CIB-nulled). These maps enter into both $\overline{T}$ and $\overline{E}$, leading to two versions of the filtered fields $\overline{\bm{X}}^{(1)}_{\ell m}$ and $\overline{\bm{X}}^{(2)}_{\ell m}$, depending on whether $T_1$ or $T_2$ is used in the filtering.

As with the asymmetric SQE $TT$ estimator (see Eq.~\ref{eq:sqe_T1T2_plus_T2T1}), we must symmetrize the GMV reconstruction:
\begin{equation}\label{eq:gmv_T1T2_plus_T2T1}
    \hat{\phi}^{GMV}_{LM} = \frac{1}{2} \left(\hat{\phi}^{GMV (1,2)}_{LM} + \hat{\phi}^{GMV (2,1)}_{LM}\right),
\end{equation}
where $\hat{\phi}^{GMV (1,2)}_{LM}$ denotes the estimator with $\overline{\bm{X}}^{(1)}_{\ell m}$ in the first leg and $\overline{\bm{X}}^{(2)}_{\ell' m'}$ in the second, and vice versa for $\hat{\phi}^{GMV (2,1)}_{LM}$. Every term of the matrix product $\overline{\bm{X}}^{T}_{\ell m} \bm{f}_{\ell \ell' L} \overline{\bm{X}}_{\ell' m'}$ for this asymmetric case is expressed explicitly in Appendix~\ref{sec:expand_gmv}.

In the $TT$-only SQE case, we decomposed the four-point function in Eq.~\ref{eq:TTTT} term-by-term, and showed how asymmetric reconstruction suppresses foreground-induced contributions. This was relatively straightforward, since only temperature maps entered and the filtering was simple. In the GMV formalism, the relevant four-point functions now involve mixtures of CMB fields $X \in \{T,E,B\}$; moreover, because of the joint filtering in Eq.~\ref{eq:xbar_expanded}, the filtered fields $\overline{T}$ and $\overline{E}$ themselves each contain contributions from both $T$ and $E$. Nonetheless, the same logic still applies: each four-point function can be decomposed into the primary lensing trispectrum, foreground trispectrum, primary bispectrum, and secondary bispectrum terms. Asymmetric reconstruction again mitigates these biases by reducing correlations between foreground residuals across the two inputs of the estimator. In practice, the resulting bias levels for the MV SQE and GMV versions of each method (standard, \mh, and cross-ILC) are nearly identical, confirming that extending the mitigation schemes to the GMV framework does not introduce additional bias or alter their relative effectiveness.

With the formalism established, we now apply them on simulations and compare the lensing biases with and without foreground mitigation.

\section{Input Maps} \label{sec:inputs}

In this work, we use \agora\ simulations \cite{agora} that include the lensed CMB and lensed non-Gaussian extragalactic foregrounds: tSZ, CIB, kSZ, and radio sources. The CIB and radio foreground maps are masked with a single-pixel point source mask with a 6.0~mJy flux cut at 150~GHz before combining with the CMB map. The tSZ map is separately masked and inpainted at the brightest peaks in the \agora\ tSZ map, with the cluster masking threshold set to a signal-to-noise ratio (SNR) $> 10$ given the SPT-3G D1 map depth. For simplicity, we restrict our analysis to idealized full-sky maps, which eliminates the complications of masking and apodization.

Because the foregrounds in the \agora\ simulations are themselves lensed, they can in principle generate a small additional contribution to the reconstructed lensing power spectrum, even in the absence of foreground-CMB correlations. Previous studies (e.g., \cite{mishra_manu_2019}) have shown that this bias is expected to be small compared to the dominant foreground-CMB coupling terms, and we do not isolate it separately in this work.

To estimate the response and the $N_{L}^{(0)}$ and $N_{L}^{(1)}$ biases of the lensing reconstruction, we use 250 full-sky Gaussian simulations. Each simulation contains an independent realization of the lensed primary CMB based on the \planck\ 2018 cosmology, along with Gaussian realizations of foregrounds whose power spectra match those of the non-Gaussian foregrounds in the \agora\ simulations.

We also add Gaussian noise to both the \agora\ and Gaussian maps. For this, we consider several scenarios. We start with noise levels matched to the SPT-3G~D1 analysis as the reference, and then test configurations with SPT-3G Full Depth noise, SPT-3G Ext-10k noise, standard-depth SO-goal noise, and extended-depth SO-goal noise. White noise levels and $1/f$ specifications are taken from previous analyses and forecasts~\cite{spt_lensing_20192020, spt3g_forecast, so_forecast, extended_so_forecast, srini_aso_correction}. Details of these configurations are provided in Section~\ref{sec:noise_levels}.

These maps are combined across the different frequency bands using ILC weights, which we describe in detail in Section~\ref{sec:ilc}. We then inverse-variance filter these simulations in harmonic space before inputting them into the quadratic estimator.

A simulation-based response is then computed from the Gaussian simulations by cross-correlating the reconstructed lensing maps with the input lensing potential and averaging over the 250 realizations. This is equivalent to the analytic expression in Eq.~\ref{eq:eq43} and is used to normalize the quadratic estimators. 
Following standard lensing reconstruction methods (e.g., \cite{act_dr6_lensing, spt_lensing_20192020}), the disconnected noise bias $N_{L}^{(0)}$ is estimated from correlations between reconstructions built from independent CMB, noise, and Gaussian foreground realizations.
The subdominant $N_{L}^{(1)}$ bias is obtained using pairs of simulations sharing the same input lensing potential but different CMB realizations. Both the $N_{L}^{(0)}$ and $N_{L}^{(1)}$ biases are subtracted out at the spectrum level.

To evaluate the residual bias in the lensing reconstruction, we apply the same reconstruction and normalization to the non-Gaussian \agora\ simulations, which include CMB and extragalactic foregrounds. The lensing bias is defined as the difference between the reconstructed and input lensing spectra after subtracting out $N_{L}^{(0),{\rm RD}}$ and $N_{L}^{(1)}$ (see Equation~\ref{eq:total_bias}), where $N_{L}^{(0),{\rm RD}}$ is the realization-dependent $N_{L}^{(0)}$ which is computed by taking correlations between the Gaussian simulations and the \agora\ realization \cite{namikawa_hanson_takahashi_2013, planck_lensing_2015}.

\subsection{Noise Levels and Experiment Configuration} \label{sec:noise_levels}

We describe the noise scenarios considered in this work below.
As the reference case, we use noise levels matching the SPT-3G D1 analysis~\cite{spt_lensing_20192020} and sample Gaussian realizations from them via \texttt{synfast} \cite{healpix}. The cross-frequency correlations are included using the method described in Appendix~A of \cite{correlated_mc_maps}.
Next, we consider a scenario with the projected noise levels for SPT-3G using data collected over seven years (2019-2023, 2025-2026) over the SPT-3G Main survey region, ``SPT-3G Full Depth". We adopt the white noise levels and $1/f$ specifications presented in \cite{spt3g_forecast} to construct the temperature and polarization noise auto-spectra for each frequency band. Then, we assume the same cross-frequency correlation levels as the SPT-3G~D1 analysis to compute the noise cross-spectra between different frequency bands. Once the full set of auto- and cross-frequency noise spectra is constructed, we generate Gaussian noise realizations following the same procedure as in the SPT-3G D1 case.

We also test the SPT-3G Ext-10k configuration, with parameters again taken from \cite{spt3g_forecast}. SPT-3G Ext-10k is a combination of the SPT-3G Main, Summer, and Wide surveys, which together expand the survey area to roughly 10,000 deg$^2$. The Summer and Wide surveys are shallower, large-area extensions of the deep 1500 deg$^2$ SPT-3G Main field. Here we continue to make the same assumptions for the cross-frequency correlations as in the SPT-3G D1 and Full Depth cases. We compute the lensing bandpowers for these three surveys individually, then combine them using inverse variance weighting, as the surveys have only small overlaps at the boundaries and are effectively disjoint \cite{spt3g_forecast, vitrier_2025}.

Then, we further include a scenario with the ``goal" noise levels of the standard SO experiment, as presented in \cite{so_forecast}. The noise curves are obtained from the publicly available SO noise model repository\footnote{\url{https://github.com/simonsobs/so_noise_models}}. Since we do not have \agora\ foreground maps at 27 and 39 GHz, we omit these two bands from this analysis,
as these channels do not contribute significantly due to their large beams suppressing the small-scale CMB modes relevant for lensing reconstruction.
Finally, we test a case with the ``goal" noise levels corresponding to the extended SO experiment. The white noise levels and $1/f$ parameters for the mid-frequency bands are taken from \cite{extended_so_forecast}. However, in \cite{srini_aso_correction}, the $1/f$ parameters for the two ultra-high-frequency bands, 225 and 280 GHz, were shown to be overly optimistic.
Therefore we instead use values from the CMB-S4 forecast for those bands, following the correction applied in \cite{srini_aso_correction}. We assume the same cross-frequency correlation structure as standard SO.

We use sky fraction values of $f_\mathrm{sky} = 0.04$, 0.04, 0.064, 0.145, 0.4, and 0.4 for SPT-3G D1, SPT-3G Main, SPT-3G Summer, SPT-3G Wide, SO Standard, and SO Extended, respectively, to scale the uncertainty of the lensing power spectrum from the full-sky simulations. These values were taken from \cite{spt3g_forecast} for SPT-3G and \cite{so_forecast} and \cite{extended_so_forecast} for SO. The experimental specifications used are summarized in Table~\ref{tab:experimental_specs} of Appendix~\ref{sec:experiment_specs}.

\subsection{tSZ-Nulled and CIB-Nulled ILC Maps} \label{sec:ilc}

\mh\ and cross-ILC both rely on foreground-cleaned ILC input maps. Here we describe how the tSZ-nulled and CIB-nulled ILC weights are produced via the constrained ILC method \cite{constrained_ilc, crossilc_2023}.

The ILC approach exploits the distinct frequency dependence of the different sky components, allowing the CMB signal to be isolated while suppressing foreground contamination. We begin by assuming the observed signal in the frequency channel $i$, in harmonic space, to be given by:
\begin{equation}\label{eq:obs}
    T^i_{\ell m} = a^i s_{\ell m} + b^i d_{\ell m} + n^i_{\ell m}
\end{equation}
where $s_{\ell m}$ is the signal we wish to isolate, $d_{\ell m}$ is a foreground component we wish to null the contribution in the final ILC reconstruction, and $n^i_{\ell m}$ accounts for all remaining contributions to the observation, including instrumental noise and other extragalactic foregrounds. The frequency dependence of $s_{\ell m}$ and $d_{\ell m}$ is encoded in the spectral energy distributions (SEDs) $\bm{a}$ and $\bm{b}$, respectively. 

The linear estimator for the reconstructed signal can be written as:
\begin{equation}\label{eq:lin_comb}
    \hat{s}_{\ell m} = \sum_i w^i_\ell T^i_{\ell m}.
\end{equation}

To ensure an unbiased response to the signal of interest, the weights $w^i_\ell$ must satisfy the condition
\begin{equation}\label{eq:ilc_cond1}
    \sum_i w^i_\ell a^i = 1.
\end{equation}
Additionally, for the resulting signal to have null response to $d_{\ell m}$, the weights must also satisfy
\begin{equation}\label{eq:ilc_cond2}
    \sum_i w^i_\ell b^i = 0.
\end{equation}

Minimizing the variance in the reconstructed signal while enforcing the constraints in Eq.~\ref{eq:ilc_cond1} and Eq.~\ref{eq:ilc_cond2} leads to the following expression for the ILC weights:
\begin{equation}
	\bm{w}_\ell = \bm{N} (\bm{A}^T \bm{C}^{-1}_\ell \bm{A})^{-1} \bm{A}^T \bm{C}^{-1}_\ell
\end{equation}
where $\bm{C}_\ell$ is the covariance matrix of the input signals at multipole $\ell$, $\bm{A} = \left[\bm{a}~\bm{b}\right]$ contains the SED vectors of the target signal and the nulled component, and $\bm{N} = \left[1~0\right]$ selects the weight vector associated with the target signal while discarding the component corresponding to the nulled SED. Note that here $\bm{C}_\ell$ refers to the covariance across frequency channels, not the $\{T,E,B\}$ covariance matrix introduced earlier in Section~\ref{sec:standard_gmv_formalism}.

The ILC approach can also be applied in the minimum-variance (MV) configuration, in which no component is deprojected. In that case, the only condition imposed on the weights construction is that of Eq. \ref{eq:ilc_cond1}, and the matrix $\bm{A}$ reduces simply to $\bm{A} = \bm{a}$.

Using the formalism described above, we generate ILC maps to be used as inputs in the lensing reconstruction. For each instrument configuration, three temperature maps are produced: MV ILC, tSZ-nulled ILC, and CIB-nulled ILC. For polarization, only the MV ILC version is constructed.

For the MV scheme, the only SED needed is that of the CMB, which is given by $\bm{a} = [1,1,\dots,1]$, since the input frequency maps are calibrated to have unit
response to the CMB. To construct the tSZ-nulled ILC, the deprojected component's SED $\bm{b}$ is set to the tSZ spectral function
\begin{equation}\label{eq:f_tsz}
    f^\nu_{\mathrm{tSZ}} = x \frac{e^x + 1}{e^x - 1} - 4 ,
\end{equation}
in which $x \equiv h\nu/k_B T_\text{CMB}$ \cite{sz_1970, sz_1972}. Here, $h$ is Planck's constant, $\nu$ the observing frequency, $k_B$ Boltzmann's constant, and $T_\text{CMB} = 2.726$ K. 

For the CIB deprojection, the CIB SED is modeled as a modified blackbody:
\begin{equation}\label{eq:f_cib}
    f^\nu_\text{CIB} \propto \nu^\beta B(\nu, T_\text{CIB})
\end{equation}
where $\beta$ is the CIB spectral index, $T_\text{CIB}$ the effective temperature of the CIB signal, and $B(\nu, T)$ the blackbody spectral function:
\begin{equation}\label{eq:bb}
    B(\nu, T) = \frac{2h\nu^3}{c^2} \frac{1}{e^{\frac{h\nu}{k_B T}}-1}.
\end{equation}
The optimal choice of $\beta$ and $T_\text{CIB}$ for CIB cleaning depends on the dataset, as the CIB arises from a mix of dusty star-forming galaxies at different redshifts. To account for this, we perform a grid search as in \cite{crossilc_2023} over parameter values $\{\beta^i, T_\text{CIB}^j\}$, with $\beta^i \in \left[1,3\right]$ and $T_\text{CIB}^j \in \left[5,35\right]$ K.
For each grid point, we perform a CIB-nulled ILC reconstruction and compute the residual CIB power in the recovered CMB maps using Agora simulations. The parameter set that minimizes the CIB residuals over the range  $\ell \in \left[0,4096\right]$ is adopted for the nulling SED $\bm{b}$. This optimization is performed independently for each experiment configuration.

To keep the analysis tractable, we assume a single modified blackbody SED for the CIB. In reality, the CIB spectrum is more complex and varies across the sky due to the mix of galaxies at different redshifts and dust temperatures. As a result, perfect nulling is not possible in practice, and some residual CIB contamination will remain even with optimized parameters.

\begin{figure}[t]
	\centering
	\includegraphics[width=\textwidth]{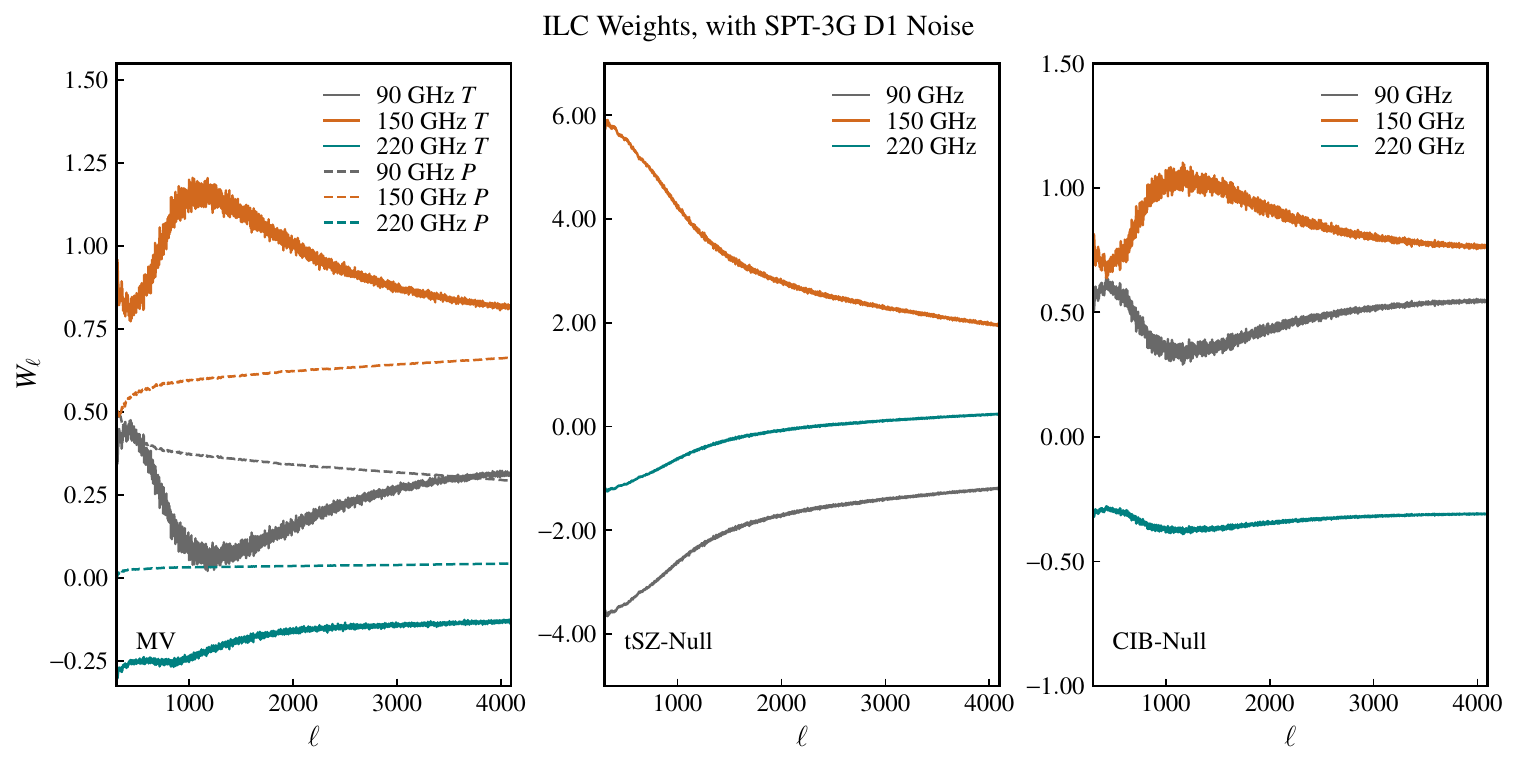}
	\centering \caption{MV, tSZ-nulled, and CIB-nulled ILC weights for SPT-3G D1 noise levels.}
	\label{fig:mv_tszn_cibn_ilc_weights_spt3g_20192020}
\end{figure}

The covariance matrix is constructed from the \agora\ inputs and the Gaussian noise models described above. For the SPT-3G reconstructions, we include the $95$, $150$, and $220\,\mathrm{GHz}$ single-frequency maps, while for the SO cases we use the $95$, $150$, $220$, and $280$ GHz maps.
Figure~\ref{fig:mv_tszn_cibn_ilc_weights_spt3g_20192020} shows the MV, tSZ-nulled, and CIB-nulled ILC weights for SPT-3G D1 noise levels.

\section{Numerical Results} \label{sec:numerical_results}

We have introduced two approaches for mitigating extragalactic foregrounds in the GMV framework: \mh\ and cross-ILC. Both methods reduce the foreground-induced lensing bias, but at the cost of increased reconstruction noise, since the foreground-nulled input maps themselves have increased variance.
To evaluate the utility of these techniques, we need to examine the trade-off between bias reduction and reconstruction noise increase. In this section, we first study the reference case with noise levels matching those of the SPT-3G D1 analysis~\cite{spt_lensing_20192020}. As the fiducial setting, we use $\ell_{\mathrm{max}}^T = 3500$ for temperature and $\ell_{\mathrm{max}}^P = 4096$ for polarization. Then in Section~\ref{sec:different_noise_levels} we extend this study to additional noise scenarios.

\subsection{Reconstruction Noise for SPT-3G D1 Noise Levels}
\label{sec:reconstruction_noise}

\begin{figure}[t]
	\centering
	\includegraphics[width=\textwidth]{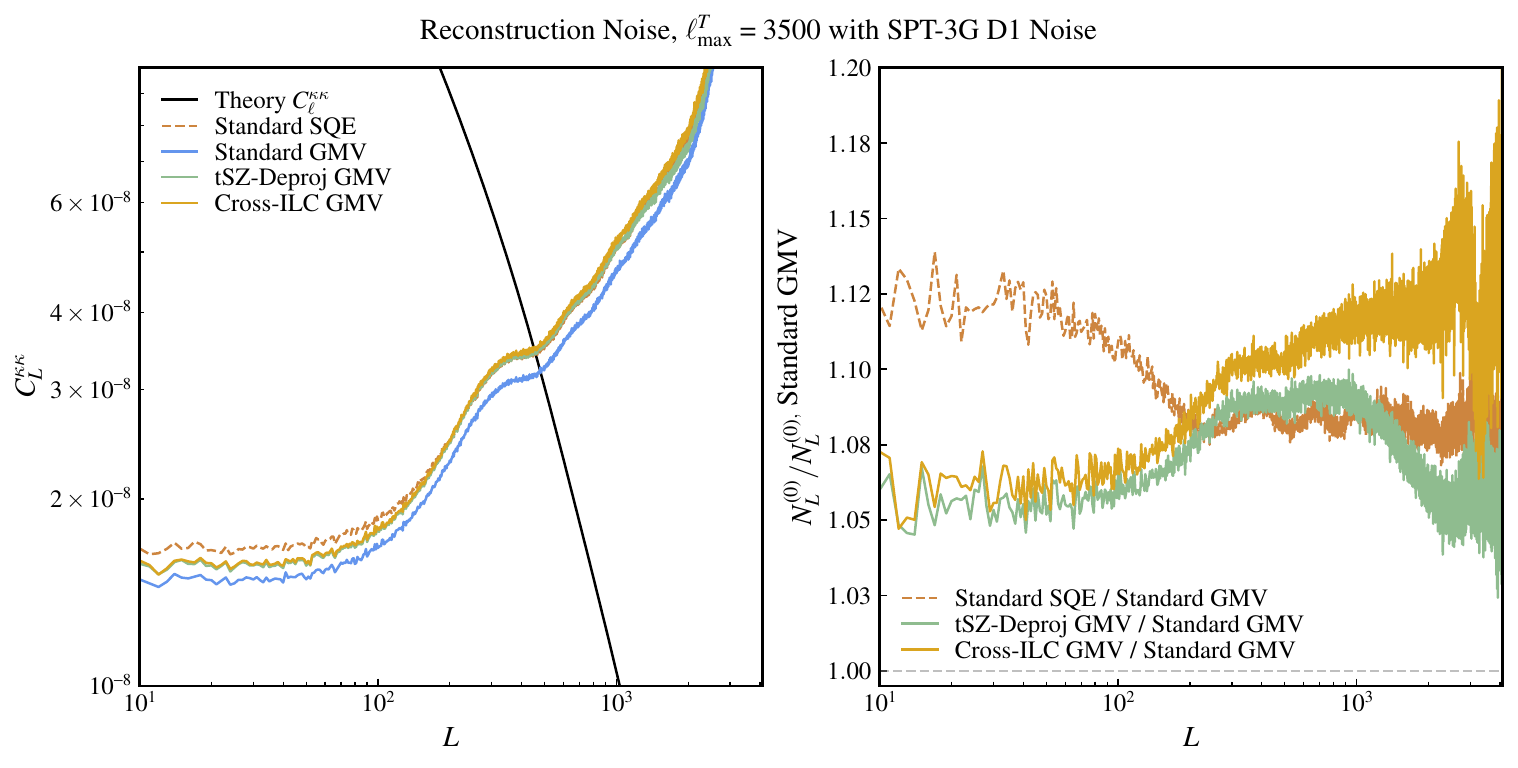}
	\centering \caption{Reconstruction noise spectra comparison, with $\ell_{\mathrm{max}}^T = 3500$.
	Left panel: Noise spectra plotted against the \camb\protect\footnotemark{} theory $\kappa$ auto-spectrum. The standard GMV case with no foreground treatment has the lowest reconstruction noise, as expected.
	Right panel: Noise spectra normalized to the standard GMV case.
	Here we see clearly the increase in noise of the standard SQE MV, \mh\ GMV, and cross-ILC GMV, with respect to standard GMV. For $L \lesssim 200$, the noise performance of foreground-mitigated GMV is better than standard SQE, leading to both gain in sensitivity and robustness against foreground biases.}
	\label{fig:n0_comparison}
\end{figure}

To numerically study the reconstruction noise for these foreground mitigation methods, we use 250 full-sky simulations with Gaussian foregrounds and noise as described in Section~\ref{sec:inputs}. The resulting reconstruction noise spectra are shown in Figure~\ref{fig:n0_comparison} for our reference case with SPT-3G D1 noise. The standard SQE MV and GMV estimators with no foreground cleaning are included for comparison against GMV with \mh\ and cross-ILC methods applied. We note that standard GMV improves the reconstruction noise by roughly 10\% over standard SQE MV, consistent with what was claimed in \cite{gmv}.

\footnotetext{\url{https://camb.info}}

Both \mh\ and cross-ILC GMV reconstructions yield higher noise than standard GMV because their foreground-nulled inputs have larger variance. The increase in noise is slightly greater for cross-ILC, where both input maps are foreground-nulled rather than just one.
For $L \gtrsim 200$, the noise penalty for \mh\ and cross-ILC is comparable to or slightly larger than that of standard SQE, at the 10-15\% level relative to standard GMV. At lower multipoles ($L \lesssim 200$), both foreground-mitigated GMV variants outperform standard SQE and differ from standard GMV by only $\sim$5\%.

While foreground mitigation introduces a 5–15\% noise penalty at fixed $\ell_{\mathrm{max}}^T = 3500$, the overall statistical precision of the reconstruction can be greatly improved by including higher temperature multipoles.
For example, increasing $\ell_{\mathrm{max}}^T$ from 3000 to 4000 reduces $N_{L}^{(0)}$ by approximately 30\% at $L = 1000$, 60\% at $L = 2000$, and 75\% at $L = 3000$.
Even the smaller step from $\ell_{\mathrm{max}}^T = 3500$ to 4000 yields improvements of 10\%, 25\%, and 60\% at the same multipoles.
These gains occur for both standard and foreground-mitigated reconstructions.
However, in the standard GMV or SQE cases, extending to such high $\ell_{\mathrm{max}}^T$ would result in statistically significant foreground biases. 
In contrast, with \mh\ and cross-ILC, higher $\ell_{\mathrm{max}}^T$ can yield substantial gains in sensitivity while remaining robust against foreground biases.

\subsection{Lensing Bias for SPT-3G D1 Noise Levels}

\begin{figure}[t]
	\centering
	\includegraphics[width=.6\textwidth]{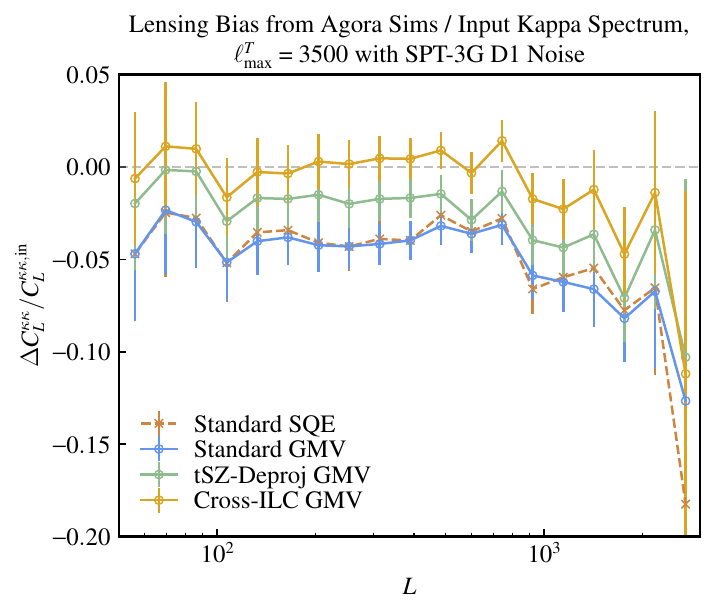}
	\centering \caption{Lensing bias comparison for $\ell_{\mathrm{max}}^T = 3500$. Here, the lensing bias is defined as $\Delta C_L^{\kappa\kappa} = C_L^{\kappa\kappa,\mathrm{recon}} - C_L^{\kappa\kappa,\mathrm{in}}$ and it is normalized by the input $\kappa$ spectrum. The error bars are measurement error for full sky. The standard SQE and GMV with no foreground treatment have comparable bias, at around 4\%. \mh\ GMV removes tSZ bias, reducing the bias to around 2\%. Cross-ILC further cleans CIB as well as tSZ, reducing the bias to be consistent with zero for $L \lesssim 1000$.}
	\label{fig:lensing_bias}
\end{figure}

In addition to the 250 Gaussian simulations, we use non-Gaussian foreground simulations from \agora\ as described in Section~\ref{sec:inputs} to test the level of reduced lensing bias from the foreground mitigation techniques. Figure~\ref{fig:lensing_bias} shows the lensing bias for each case at the fiducial $\ell_{\mathrm{max}}^T = 3500$, expressed as ratios with the input $\kappa$ spectrum in the denominator. The error bars are measurement error for the idealized full sky experiment.

Here, the lensing bias for $L \lesssim 1000$ is around 4\% for standard SQE MV and GMV, 2\% for \mh\ GMV, and negligible for cross-ILC GMV. Because these values all correspond to MV reconstructions in which all estimator pairs contribute, the bias is lower than what would be expected for a $TT$-only SQE reconstruction.
The level of residual bias in the cross-ILC case depends on how effectively the CIB is suppressed by the ILC weights. As discussed in Section~\ref{sec:ilc}, perfect CIB nulling is not achievable in practice because the true CIB SED is spatially variable and deviates from a single modified blackbody model. Nevertheless, in this configuration the residual CIB is sufficiently suppressed to reduce the cross-ILC GMV bias to a level consistent with zero for most of the $L$ range.
Overall, both the \mh\ and cross-ILC implementations perform as intended, substantially reducing the lensing bias.

\begin{figure}[t]
	\centering
	\includegraphics[width=.6\textwidth]{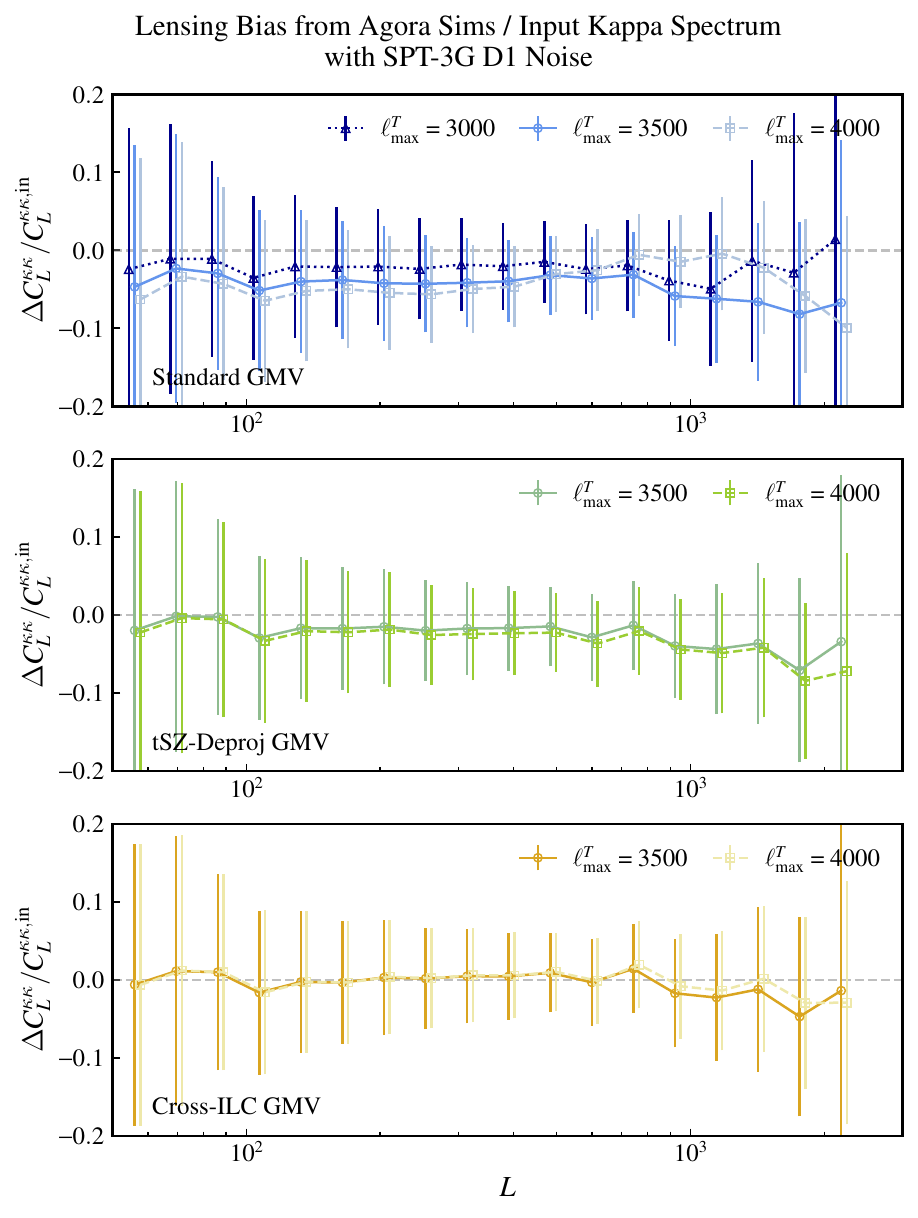}
	\centering \caption{Lensing bias comparison for $\ell_{\mathrm{max}}^T = 3000, 3500, 4000$ cases. The error bars correspond to measurement errors for a full sky reconstruction. The data points for the different $\ell_{\mathrm{max}}^T$ values are shifted horizontally for clarity.
    We see that going up in $\ell_{\mathrm{max}}^T$ leads to a significant increase in lensing bias in the standard case with no foreground mitigation, but the penalty is significantly reduced with foreground mitigation applied.}
	\label{fig:lensing_bias_different_lmaxT}
\end{figure}

We also experiment with varying $\ell_{\mathrm{max}}^T$ to test how much the addition of high-$\ell$ temperature modes amplifies the lensing bias. As shown in Figure~\ref{fig:lensing_bias_different_lmaxT}, for $L < 500$, increasing $\ell_{\mathrm{max}}^T$ in the standard GMV case consistently raises the lensing bias, from $\sim2\%$ at $\ell_{\mathrm{max}}^T = 3000$ to $\sim5\%$ at $\ell_{\mathrm{max}}^T = 4000$. In contrast, the \mh\ and cross-ILC GMV reconstructions remain largely stable: for \mh\ the change between $\ell_{\mathrm{max}}^T = 3500$ and $4000$ is only about 0.5\%, and for cross-ILC the shift is essentially zero. Since extending $\ell_{\mathrm{max}}^T$ greatly lowers reconstruction noise as discussed in Section~\ref{sec:reconstruction_noise}, the ability to do so without introducing additional lensing bias is a key advantage of these foreground-mitigated estimators.

The behavior at $L > 500$ for the $\ell_{\mathrm{max}}^T = 4000$ standard GMV is less straightforward: the lensing bias exhibits an anomalous ``bump'' that peaks at around $L = 1000$. This feature is absent when including only tSZ or only CIB in the \agora\ simulations, but reappears when both are present, strongly suggesting that it originates from tSZ–CIB correlations. We provide further discussion of this effect in Appendix~\ref{sec:tszxcib}. 
This behavior is not observed in foreground-mitigated reconstructions.

As we will show in Section~\ref{sec:different_noise_levels}, for an analysis with $\ell_{\mathrm{max}}^T = 3500$ and noise levels comparable to the SPT-3G D1 dataset, the mean lensing bias in the range $50 < L < 1000$ is at the level of 0.45$\sigma$ without foreground mitigation, where $\sigma$ denotes the statistical uncertainty on the lensing amplitude. Applying the \mh\ GMV estimator reduces this bias to $0.2\sigma$, while the cross-ILC GMV estimator further suppresses it to $0.08\sigma$. These results demonstrate that foreground mitigation reduces the residual bias to well below the statistical error for realistic experiment configurations. The quantitative comparison between lensing bias and uncertainty across different noise scenarios is shown in Figure~\ref{fig:lensing_bias_vs_uncertainty_different_noise} and discussed in more detail in Section~\ref{sec:different_noise_levels}.

\subsection{Testing Different Noise Levels} \label{sec:different_noise_levels}

\begin{figure}[t]
	\centering
	\includegraphics[width=\textwidth]{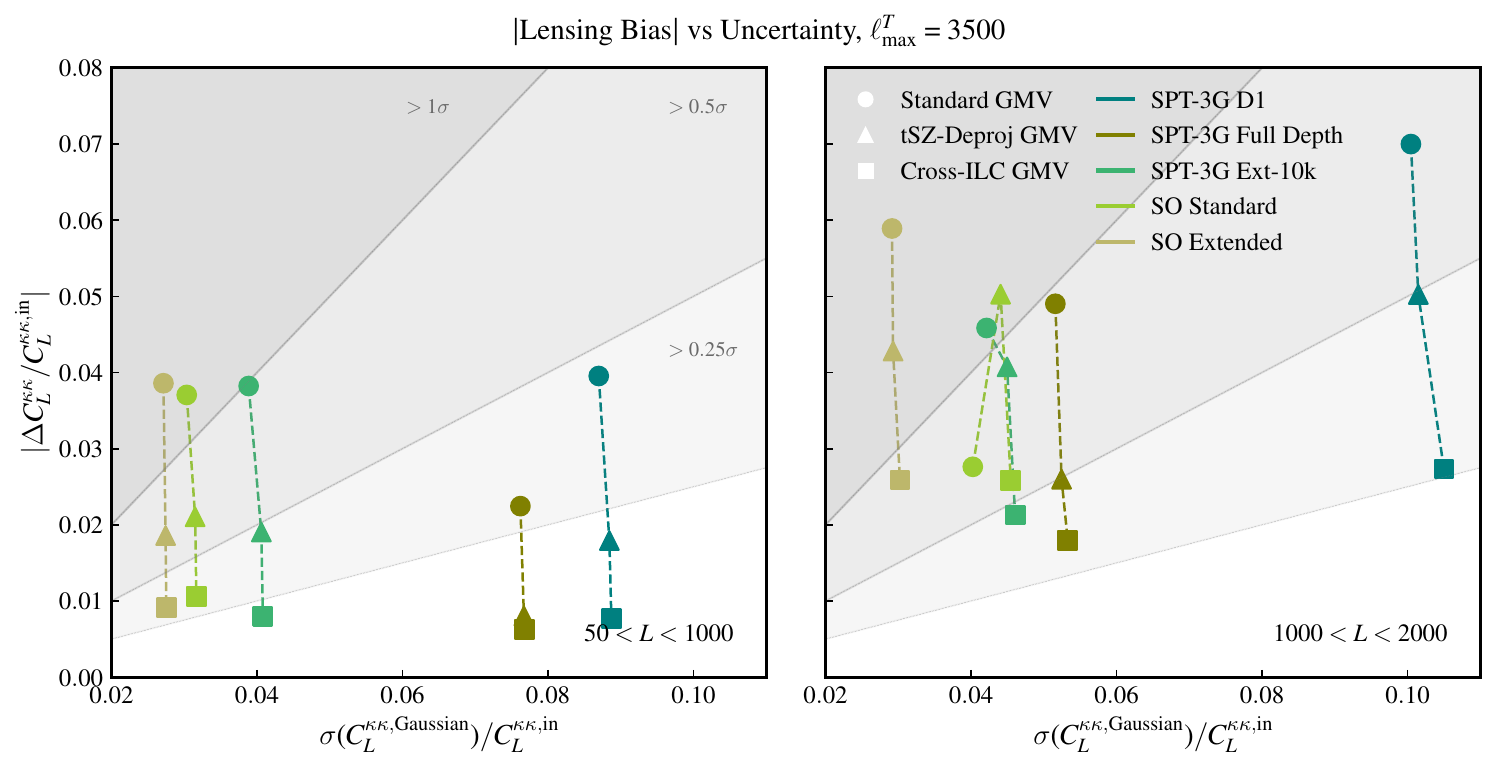}
	\centering \caption{Lensing bias comparison as a function of uncertainty, at $\ell_{\mathrm{max}}^T = 3500$, for different noise scenarios. The different colors represent different noise cases, and marker shapes represent reconstruction type. The values are averaged (standard average, not inverse variance weighted) for two chosen L bins: $50 < L < 1000$ and $1000 < L < 2000$. The uncertainty values are the standard deviation of the sims, scaled by the $1/\sqrt{f_\mathrm{sky}}$ for the corresponding experiment.}
	\label{fig:lensing_bias_vs_uncertainty_different_noise}
\end{figure}

In addition to the reference case using SPT-3G D1 noise levels, we also consider noise scenarios for SPT-3G Full Depth, SPT-3G Ext-10k, SO Standard, and SO Extended experiments. Figure~\ref{fig:lensing_bias_vs_uncertainty_different_noise} shows the lensing bias as a function of statistical uncertainty on $C_L^{\phi\phi}$ for each of these experiments, with $\ell_{\mathrm{max}}^T = 3500$. The uncertainties correspond to the standard deviation of 250 Gaussian simulations, scaled by $1/\sqrt{f_\mathrm{sky}}$. We show results for two $L$ bins, with values averaged within each bin. The shaded regions indicate thresholds where the lensing bias exceeds 0.25, 0.5, or 1 times the statistical uncertainty. The darkest gray region corresponds to $|\textnormal{lensing bias}| > 1\sigma$, where the bias dominates over statistical uncertainty, while the lighter areas indicate smaller, subdominant biases.

The relative ordering of uncertainties is consistent across both bins: largest for SPT-3G D1, followed by SPT-3G Full Depth, then SPT-3G Ext-10k and SO Standard, with the lowest uncertainties for SO Extended. Despite the low instrumental noise levels for SPT-3G, the larger sky fraction of SO leads to lower overall uncertainties after the $f_\textnormal{sky}$ correction.
The increase in uncertainty from applying foreground mitigation is very small in all experiments, especially in the $50 < L < 1000$ bin. This minor loss in precision is far outweighed by the reduction in reconstruction noise gained by extending $\ell_{\mathrm{max}}^T$, as discussed in Section~\ref{sec:reconstruction_noise}.

In the lower multipole range ($50 < L < 1000$), the three reconstruction methods yield fairly consistent lensing biases across the different noise scenarios: around 4\% for standard GMV, 2\% for \mh, and $< 1\%$ for cross-ILC.
We see that for large-area surveys such as SPT-3G Ext-10k and SO, the standard GMV estimator produces a lensing bias comparable to, or exceeding, the $1\sigma$ statistical uncertainty on the lensing amplitude. Applying foreground mitigation techniques can significantly reduce this bias to below 0.5$\sigma$, ensuring that residual foreground contamination remains subdominant to statistical errors.

We note that for the SPT-3G Full Depth configuration, which has the lowest experimental noise among all cases considered, the \mh\ GMV estimator yields a residual bias comparable to that of cross-ILC. When the ILC is less dominated by noise, the optimal weights are able to suppress foreground power more aggressively in both the MV and tSZ-nulled inputs. In particular, the residual CIB power in the tSZ-nulled input is much lower in SPT-3G Full Depth than in higher-noise experiments. As a result, both inputs of the \mh\ estimator are more CMB-dominated, making \mh\ alone nearly as effective as cross-ILC.

At higher multipoles ($1000 < L < 2000$ bin), the weight of the $TT$ estimator increases in the MV combination, leading to a larger lensing bias compared to the $50 < L < 1000$ range. Without foreground mitigation, the bias reaches up to 7\% (SPT-3G D1 noise), but with foreground mitigation this can be reduced to $< 3\%$.
In this higher $L$ regime, the SO Standard case shows the same anomalous feature noted previously in Figure~\ref{fig:lensing_bias_different_lmaxT}, where the standard GMV reconstruction exhibits an unexpected dip in bias. We attribute this to tSZ-CIB correlations, as discussed further in Appendix~\ref{sec:tszxcib}.

While our comparisons span a range of experiments with differing noise levels, frequency coverage, and angular resolution, it is useful to briefly isolate the role of beam size alone. Higher angular resolution (i.e., smaller beam FWHM) increases the relative weights of smaller-scale temperature modes, which increases the lensing signal-to-noise in temperature-dominated reconstructions. However, these small-scale modes are also where extragalactic foregrounds such as tSZ and CIB have significant power. Thus, higher-resolution experiments gain additional statistical sensitivity but are simultaneously more susceptible to foreground-induced biases at high multipoles. This trade-off again highlights the importance of robust foreground mitigation techniques, such as \mh\ and cross-ILC, to enable the use of small-scale temperature information while maintaining control over biases.

For upcoming large-area surveys such as SPT-3G Ext-10k and SO, even cross-ILC leaves a residual bias of $\gtrsim 0.5\sigma$ in the $1000 < L < 2000$ range at $\ell_{\mathrm{max}}^T = 3500$.
This suggests that temperature-based reconstructions at high multipoles will be limited by foreground contamination even after applying \mh\ and cross-ILC methods.
Further bias reduction will require lowering $\ell_{\mathrm{max}}^T$, using the shear estimator~\cite{schaan_2019, qu_2023, carron_2024}, using combinations of different foreground mitigation methods~\cite{darwish, sailor}, or relying more heavily on polarization estimators in this regime.

\section{Conclusion} \label{sec:conclusion}

In this work, we address a central challenge for high-precision CMB lensing reconstruction: biases from extragalactic foregrounds, particularly the tSZ effect and the CIB. As current and upcoming surveys such as SPT-3G and SO push toward percent-level constraints on the lensing power spectrum, it is essential to develop methods that suppress these biases to below the statistical uncertainty.

We present the implementations of \mh\ and cross-ILC within the GMV quadratic estimator framework. Both approaches mitigate foreground biases via asymmetric lensing reconstruction, where correlations between foreground residuals in the two inputs of the estimator are reduced. Using a detailed term-by-term analysis of the four-point function, we show how these methods suppress both the bispectrum and trispectrum contributions induced by foreground non-Gaussianity.

We then evaluate their performance using simulations spanning a range of experimental noise scenarios and $\ell_{\mathrm{max}}^T$ values. For the SPT-3G D1 reference case with $\ell_{\mathrm{max}}^T = 3500$, the standard GMV reconstruction exhibits a lensing bias of around 4\%. \mh\ reduces this to roughly 2\%, while cross-ILC suppresses it to nearly zero. When varying $\ell_{\mathrm{max}}^T$, the standard GMV lensing bias increases by $\sim3\%$ when going from $\ell_{\mathrm{max}}^T = 3000$ to $4000$, whereas the \mh\ and cross-ILC cases remain essentially unchanged at the sub-percent level. These results show that foreground mitigation reduces the lensing bias as intended, and allows high-$\ell$ temperature modes to be used safely without introducing additional bias.

Although foreground mitigation increases the reconstruction noise by 5–15\% at fixed $\ell_{\mathrm{max}}^T = 3500$, this cost is outweighed by the much larger gain in statistical precision achieved when higher temperature multipoles are included. Increasing $\ell_{\mathrm{max}}^T$ from 3000 to 4000 yields up to a 75\% reduction in reconstruction noise, and up to 60\% for the smaller step from $\ell_{\mathrm{max}}^T = 3500$ to 4000. In the absence of foreground treatment, these gains are offset by rapidly increasing foreground bias. However, with \mh\ and cross-ILC, the same improvement in precision is achieved with only minimal additional bias. Thus, foreground-mitigated GMV estimators enable both increased sensitivity and greater robustness to foreground contamination in future experiments.

These methods are particularly valuable for upcoming surveys like SPT-3G Ext-10k, where temperature-dominated reconstructions would otherwise be limited by foreground biases, and for the extended SO, where controlling temperature foregrounds remains important despite the strong polarization sensitivity. However, at the precision of these future surveys, residual biases of $> 0.25\sigma$ for $L < 1000$ and $> 0.5\sigma$ for $1000 < L < 2000$ can remain even with \mh\ or cross-ILC. Further improvement will be necessary to keep foreground systematics subdominant compared to the statistical uncertainty.

Finally, while our results are based on idealized simulations, they demonstrate that robust temperature-based lensing reconstruction with controlled foreground bias is achievable for upcoming surveys. Real-world analyses will require addressing additional complexities such as survey masks (which correlate with the foreground fields themselves), spatially varying noise, and other sources of mode coupling that can introduce secondary biases. Incorporating these effects will allow \mh\ and cross-ILC to be deployed and validated on data at full scale. With these extensions, the methods presented here provide a clear path toward unlocking the full statistical power of next-generation CMB lensing datasets.

\appendix

\section{Terms of the GMV Estimator with Asymmetric Temperature Inputs} \label{sec:expand_gmv}

In this section, we expand the matrix product $\overline{\bm{X}}^{T}_{\ell m} \bm{f}_{\ell \ell' L} \overline{\bm{X}}_{\ell' m'}$ in the case of asymmetric temperature inputs $T_1 \neq T_2$, and explicitly write out the symmetrized terms of Eq.~\ref{eq:gmv_T1T2_plus_T2T1}. We omit the normalization factor $\frac{\lambda(L)}{2}$, the summations over multipoles, the $(-1)^M$ factor, and the Wigner-3j symbol of Eq.~\ref{eq:eq52}, and focus solely on the matrix contraction. Below are the complete 16 combinations of the inverse-variance weighted fields:
\begin{align}\label{eq:configuration_T1T2}
	&\overline{T}_1\overline{T}_2:\nonumber\\
    &\frac{1}{2}\frac{C_{\ell_1}^{EE}T_1 - C_{\ell_1}^{T_1E}E}{D_{\ell_1}^{1}} f^{TT} \frac{C_{\ell_2}^{EE}T_2 - C_{\ell_2}^{T_2E}E}{D_{\ell_2}^{2}}\nonumber\\
	&\hspace{0.5cm}=\frac{C_{\ell_1}^{EE}C_{\ell_2}^{EE}}{2D_{\ell_1}^{1}D_{\ell_2}^{2}} f^{TT} T_1T_2 - \frac{C_{\ell_1}^{EE}C_{\ell_2}^{T_2E}}{2D_{\ell_1}^{1}D_{\ell_2}^{2}} f^{TT} T_1E
	- \frac{C_{\ell_1}^{T_1E}C_{\ell_2}^{EE}}{2D_{\ell_1}^{1}D_{\ell_2}^{2}} f^{TT} ET_2 + \frac{C_{\ell_1}^{T_1E}C_{\ell_2}^{T_2E}}{2D_{\ell_1}^{1}D_{\ell_2}^{2}} f^{TT} EE,
\end{align}
\begin{align}\label{eq:configuration_T2T1}
	&\overline{T}_2\overline{T}_1:\nonumber\\
	&\frac{1}{2}\frac{C_{\ell_1}^{EE}T_2 - C_{\ell_1}^{T_2E}E}{D_{\ell_1}^{2}} f^{TT} \frac{C_{\ell_2}^{EE}T_1 - C_{\ell_2}^{T_1E}E}{D_{\ell_2}^{1}}\nonumber\\
	&\hspace{0.5cm}=\frac{C_{\ell_1}^{EE}C_{\ell_2}^{EE}}{2D_{\ell_1}^{2}D_{\ell_2}^{1}} f^{TT} T_2T_1 - \frac{C_{\ell_1}^{EE}C_{\ell_2}^{T_1E}}{2D_{\ell_1}^{2}D_{\ell_2}^{1}} f^{TT} T_2E
	- \frac{C_{\ell_1}^{T_2E}C_{\ell_2}^{EE}}{{2D_{\ell_1}^{2}D_{\ell_2}^{1}}} f^{TT} ET_1 + \frac{C_{\ell_1}^{T_2E}C_{\ell_2}^{T_1E}}{2D_{\ell_1}^{2}D_{\ell_2}^{1}} f^{TT} EE,
\end{align}
\begin{align}\label{eq:configuration_E1E2}
	&\overline{E}_1\overline{E}_2:\nonumber\\
	&\frac{1}{2}\frac{C_{\ell_1}^{T_1T_1}E - C_{\ell_1}^{T_1E}T_1}{D_{\ell_1}^{1}} f^{EE} \frac{C_{\ell_2}^{T_2T_2}E - C_{\ell_2}^{T_2E}T_2}{D_{\ell_2}^{2}}\nonumber\\
	&\hspace{0.5cm}=\frac{C_{\ell_1}^{T_1T_1}C_{\ell_2}^{T_2T_2}}{2D_{\ell_1}^{1}D_{\ell_2}^{2}} f^{EE} EE - \frac{C_{\ell_1}^{T_1T_1}C_{\ell_2}^{T_2E}}{2D_{\ell_1}^{1}D_{\ell_2}^{2}} f^{EE} ET_2
	- \frac{C_{\ell_1}^{T_1E}C_{\ell_2}^{T_2T_2}}{2D_{\ell_1}^{1}D_{\ell_2}^{2}} f^{EE} T_1E + \frac{C_{\ell_1}^{T_1E}C_{\ell_2}^{T_2E}}{2D_{\ell_1}^{1}D_{\ell_2}^{2}} f^{EE} T_1T_2,
\end{align}
\begin{align}\label{eq:configuration_E2E1}
	&\overline{E}_2\overline{E}_1:\nonumber\\
	&\frac{1}{2}\frac{C_{\ell_1}^{T_2T_2}E - C_{\ell_1}^{T_2E}T_2}{D_{\ell_1}^{2}} f^{EE} \frac{C_{\ell_2}^{T_1T_1}E - C_{\ell_2}^{T_1E}T_1}{D_{\ell_2}^{1}}\nonumber\\
	&\hspace{0.5cm}=\frac{C_{\ell_1}^{T_2T_2}C_{\ell_2}^{T_1T_1}}{2D_{\ell_1}^{2}D_{\ell_2}^{1}} f^{EE} EE - \frac{C_{\ell_1}^{T_2T_2}C_{\ell_2}^{T_1E}}{2D_{\ell_1}^{2}D_{\ell_2}^{1}} f^{EE} ET_1
	- \frac{C_{\ell_1}^{T_2E}C_{\ell_2}^{T_1T_1}}{2D_{\ell_1}^{2}D_{\ell_2}^{1}} f^{EE} T_2E + \frac{C_{\ell_1}^{T_2E}C_{\ell_2}^{T_1E}}{2D_{\ell_1}^{2}D_{\ell_2}^{1}} f^{EE} T_2T_1,
\end{align}
\begin{align}\label{eq:configuration_T1E2}
	&\overline{T}_1\overline{E}_2:\nonumber\\
	&\frac{1}{2}\frac{C_{\ell_1}^{EE}T_1 - C_{\ell_1}^{T_1E}E}{D_{\ell_1}^{1}} f^{TE} \frac{C_{\ell_2}^{T_2T_2}E - C_{\ell_2}^{T_2E}T_2}{D_{\ell_2}^{2}}\nonumber\\
	&\hspace{0.5cm}=\frac{C_{\ell_1}^{EE}C_{\ell_2}^{T_2T_2}}{2D_{\ell_1}^{1}D_{\ell_2}^{2}} f^{TE} T_1E - \frac{C_{\ell_1}^{EE}C_{\ell_2}^{T_2E}}{2D_{\ell_1}^{1}D_{\ell_2}^{2}} f^{TE} T_1T_2
	- \frac{C_{\ell_1}^{T_1E}C_{\ell_2}^{T_2T_2}}{2D_{\ell_1}^{1}D_{\ell_2}^{2}} f^{TE} EE + \frac{C_{\ell_1}^{T_1E}C_{\ell_2}^{T_2E}}{2D_{\ell_1}^{1}D_{\ell_2}^{2}} f^{TE} ET_2,
\end{align}
\begin{align}\label{eq:configuration_T2E1}
	&\overline{T}_2\overline{E}_1:\nonumber\\
	&\frac{1}{2}\frac{C_{\ell_1}^{EE}T_2 - C_{\ell_1}^{T_2E}E}{D_{\ell_1}^{2}} f^{TE} \frac{C_{\ell_2}^{T_1T_1}E - C_{\ell_2}^{T_1E}T_1}{D_{\ell_2}^{1}}\nonumber\\
	&\hspace{0.5cm}=\frac{C_{\ell_1}^{EE}C_{\ell_2}^{T_1T_1}}{2D_{\ell_1}^{2}D_{\ell_2}^{1}} f^{TE} T_2E - \frac{C_{\ell_1}^{EE}C_{\ell_2}^{T_1E}}{2D_{\ell_1}^{2}D_{\ell_2}^{1}} f^{TE} T_2T_1
	- \frac{C_{\ell_1}^{T_2E}C_{\ell_2}^{T_1T_1}}{2D_{\ell_1}^{2}D_{\ell_2}^{1}} f^{TE} EE + \frac{C_{\ell_1}^{T_2E}C_{\ell_2}^{T_1E}}{2D_{\ell_1}^{2}D_{\ell_2}^{1}} f^{TE} ET_1,
\end{align}
\begin{align}\label{eq:configuration_E1T2}
	&\overline{E}_1\overline{T}_2:\nonumber\\
	&\frac{1}{2}\frac{C_{\ell_1}^{T_1T_1}E - C_{\ell_1}^{T_1E}T_1}{D_{\ell_1}^{1}} f^{TE} \frac{C_{\ell_2}^{EE}T_2 - C_{\ell_2}^{T_2E}E}{D_{\ell_2}^{2}}\nonumber\\
	&\hspace{0.5cm}=\frac{C_{\ell_1}^{T_1T_1}C_{\ell_2}^{EE}}{2D_{\ell_1}^{1}D_{\ell_2}^{2}} f^{TE} ET_2 - \frac{C_{\ell_1}^{T_1T_1}C_{\ell_2}^{T_2E}}{2D_{\ell_1}^{1}D_{\ell_2}^{2}} f^{TE} EE
	- \frac{C_{\ell_1}^{T_1E}C_{\ell_2}^{EE}}{2D_{\ell_1}^{1}D_{\ell_2}^{2}} f^{TE} T_1T_2 + \frac{C_{\ell_1}^{T_1E}C_{\ell_2}^{T_2E}}{2D_{\ell_1}^{1}D_{\ell_2}^{2}} f^{TE} T_1E,
\end{align}
\begin{align}\label{eq:configuration_E2T1}
	&\overline{E}_2\overline{T}_1:\nonumber\\
	&\frac{1}{2}\frac{C_{\ell_1}^{T_2T_2}E - C_{\ell_1}^{T_2E}T_2}{D_{\ell_1}^{2}} f^{TE} \frac{C_{\ell_2}^{EE}T_1 - C_{\ell_2}^{T_1E}E}{D_{\ell_2}^{1}}\nonumber\\
	&\hspace{0.5cm}=\frac{C_{\ell_1}^{T_2T_2}C_{\ell_2}^{EE}}{2D_{\ell_1}^{2}D_{\ell_2}^{1}} f^{TE} ET_1 - \frac{C_{\ell_1}^{T_2T_2}C_{\ell_2}^{T_1E}}{2D_{\ell_1}^{2}D_{\ell_2}^{1}} f^{TE} EE
	- \frac{C_{\ell_1}^{T_2E}C_{\ell_2}^{EE}}{2D_{\ell_1}^{2}D_{\ell_2}^{1}} f^{TE} T_2T_1 + \frac{C_{\ell_1}^{T_2E}C_{\ell_2}^{T_1E}}{2D_{\ell_1}^{2}D_{\ell_2}^{1}} f^{TE} T_2E,
\end{align}
\begin{align}\label{eq:configuration_T1B}
	&\overline{T}_1\overline{B}:\nonumber\\
	&\frac{1}{2}\frac{C_{\ell_1}^{EE}T_1 - C_{\ell_1}^{T_1E}E}{D_{\ell_1}^{1}} f^{TB} \frac{B}{C_{\ell_2}^{BB}}
	=\frac{C_{\ell_1}^{EE}}{2D_{\ell_1}^{1}C_{\ell_2}^{BB}} f^{TB} T_1B - \frac{C_{\ell_1}^{T_1E}}{2D_{\ell_1}^{1}C_{\ell_2}^{BB}} f^{TB} EB,
\end{align}
\begin{align}\label{eq:configuration_T2B}
	&\overline{T}_2\overline{B}:\nonumber\\
	&\frac{1}{2}\frac{C_{\ell_1}^{EE}T_2 - C_{\ell_1}^{T_2E}E}{D_{\ell_1}^{2}} f^{TB} \frac{B}{C_{\ell_2}^{BB}}
	=\frac{C_{\ell_1}^{EE}}{2D_{\ell_1}^{2}C_{\ell_2}^{BB}} f^{TB} T_2B - \frac{C_{\ell_1}^{T_2E}}{2D_{\ell_1}^{2}C_{\ell_2}^{BB}} f^{TB} EB,
\end{align}
\begin{align}\label{eq:configuration_BT2}
	&\overline{B}\overline{T}_2:\nonumber\\
	&\frac{1}{2}\frac{B}{C_{\ell_1}^{BB}} f^{TB} \frac{C_{\ell_2}^{EE}T_2 - C_{\ell_2}^{T_2E}E}{D_{\ell_2}^{2}}
	=\frac{C_{\ell_2}^{EE}}{2C_{\ell_1}^{BB}D_{\ell_2}^{2}} f^{TB} BT_2 - \frac{C_{\ell_2}^{T_2E}}{2C_{\ell_1}^{BB}D_{\ell_2}^{2}} f^{TB} BE,
\end{align}
\begin{align}\label{eq:configuration_BT1}
	&\overline{B}\overline{T}_1:\nonumber\\
	&\frac{1}{2}\frac{B}{C_{\ell_1}^{BB}} f^{TB} \frac{C_{\ell_2}^{EE}T_1 - C_{\ell_2}^{T_1E}E}{D_{\ell_2}^{1}}
	=\frac{C_{\ell_2}^{EE}}{2C_{\ell_1}^{BB}D_{\ell_2}^{1}} f^{TB} BT_1 - \frac{C_{\ell_2}^{T_1E}}{2C_{\ell_1}^{BB}D_{\ell_2}^{1}} f^{TB} BE,
\end{align}
\begin{align}\label{eq:configuration_E1B}
	&\overline{E}_1\overline{B}:\nonumber\\
	&\frac{1}{2}\frac{C_{\ell_1}^{T_1T_1}E - C_{\ell_1}^{T_1E}T_1}{D_{\ell_1}^{1}} f^{EB} \frac{B}{C_{\ell_2}^{BB}}
	=\frac{C_{\ell_1}^{T_1T_1}}{2D_{\ell_1}^{1}C_{\ell_2}^{BB}} f^{EB} EB - \frac{C_{\ell_1}^{T_1E}}{2D_{\ell_1}^{1}C_{\ell_2}^{BB}} f^{EB} T_1B,
\end{align}
\begin{align}\label{eq:configuration_E2B}
	&\overline{E}_2\overline{B}:\nonumber\\
	&\frac{1}{2}\frac{C_{\ell_1}^{T_2T_2}E - C_{\ell_1}^{T_2E}T_2}{D_{\ell_1}^{2}} f^{EB} \frac{B}{C_{\ell_2}^{BB}}
	=\frac{C_{\ell_1}^{T_2T_2}}{2D_{\ell_1}^{2}C_{\ell_2}^{BB}} f^{EB} EB - \frac{C_{\ell_1}^{T_2E}}{2D_{\ell_1}^{2}C_{\ell_2}^{BB}} f^{EB} T_2B,
\end{align}
\begin{align}\label{eq:configuration_BE2}
	&\overline{B}\overline{E}_2:\nonumber\\
	&\frac{1}{2}\frac{B}{C_{\ell_1}^{BB}} f^{EB} \frac{C_{\ell_2}^{T_2T_2}E - C_{\ell_2}^{T_2E}T_2}{D_{\ell_2}^{2}}
	=\frac{C_{\ell_2}^{T_2T_2}}{2C_{\ell_1}^{BB}D_{\ell_2}^{2}} f^{EB} BE - \frac{C_{\ell_2}^{T_2E}}{2C_{\ell_1}^{BB}D_{\ell_2}^{2}} f^{EB} BT_2,
\end{align}
\begin{align}\label{eq:configuration_BE1}
	&\overline{B}\overline{E}_1:\nonumber\\
	&\frac{1}{2}\frac{B}{C_{\ell_1}^{BB}} f^{EB} \frac{C_{\ell_2}^{T_1T_1}E - C_{\ell_2}^{T_1E}T_1}{D_{\ell_2}^{1}}
	=\frac{C_{\ell_2}^{T_1T_1}}{2C_{\ell_1}^{BB}D_{\ell_2}^{1}} f^{EB} BE - \frac{C_{\ell_2}^{T_1E}}{2C_{\ell_1}^{BB}D_{\ell_2}^{1}} f^{EB} BT_1.
\end{align}
Note that these include the factor of $\frac{1}{2}$ from symmetrizing. Also note that since tSZ and CIB are not polarized, all $TE$ correlations only come from $\Lambda$CDM CMB, so we may assume $C_{\ell}^{T_1E} = C_{\ell}^{T_2E} = C_{\ell}^{ET_1} = C_{\ell}^{ET_2}$.

\section{Summary of Experimental Specifications} \label{sec:experiment_specs}

\begin{table}
    \footnotesize
    \begin{tabular}{l | c | c | c | c }
    \hline\hline
    Band & 95 GHz & 150 GHz & 220 GHz & 280 GHz \\
    \hline\hline
    \multicolumn{5}{l}{SPT-3G D1:} \\
    \hline
    $f_\mathrm{sky}$ & \multicolumn{4}{c}{0.04} \\ \hline
    $\sigma$ [arcmin] & 1.6 & 1.2 & 1.0 & \\ \hline
    $\Delta_{T}$ [$\mu$K-arcmin] & 5.5 & 4.5 & 16.4 & \\ \hline
    $\ell_{\textnormal{knee},T}$ & 1200 & 1900 & 2100 & \\ \hline
    $\alpha_{T}$ & -4.2 & -4.1 & -3.9 & \\ \hline
    $\ell_{\textnormal{knee},P}$ & \multicolumn{4}{c}{300} \\ \hline
    $\alpha_{P}$ & \multicolumn{4}{c}{-1} \\ \hline\hline
    \multicolumn{5}{l}{SPT-3G Main:} \\
    \hline
    $f_\mathrm{sky}$ & \multicolumn{4}{c}{0.04} \\ \hline
    $\sigma$ [arcmin] & 1.6 & 1.2 & 1.0 & \\ \hline
    $\Delta_{T}$ [$\mu$K-arcmin] & 2.5 & 2.1 & 7.6 & \\ \hline
    $\ell_{\textnormal{knee},T}$ & 1200 & 2200 & 2300 & \\ \hline
    $\alpha_{T}$ & -3 & -4 & -4 & \\ \hline
    $\ell_{\textnormal{knee},P}$ & \multicolumn{4}{c}{300} \\ \hline
    $\alpha_{P}$ & \multicolumn{4}{c}{-1} \\ \hline\hline
    \multicolumn{5}{l}{SPT-3G Summer:} \\
    \hline
    $f_\mathrm{sky}$ & \multicolumn{4}{c}{0.064} \\ \hline
    $\sigma$ [arcmin] & 1.6 & 1.2 & 1.0 & \\ \hline
    $\Delta_{T}$ [$\mu$K-arcmin] & 9.7 & 9.1 & 29.1 & \\ \hline
    $\ell_{\textnormal{knee},T}$ & 1600 & 2600 & 2600 \\ \hline
    $\alpha_{T}$ & -4.5 & -4 & -3.9 \\ \hline
    $\ell_{\textnormal{knee},P}$ & 300 & 490 & 500 \\ \hline
    $\alpha_{P}$ & -2.2 & -2 & -2.5 & \\ \hline\hline
    \multicolumn{5}{l}{SPT-3G Wide:} \\
    \hline
    $f_\mathrm{sky}$ & \multicolumn{4}{c}{0.145} \\ \hline
    $\sigma$ [arcmin] & 1.6 & 1.2 & 1.0 & \\ \hline
    $\Delta_{T}$ [$\mu$K-arcmin] & 14 & 12 & 42 & \\ \hline
    $\ell_{\textnormal{knee},T}$ & 1600 & 2600 & 2600 & \\ \hline
    $\alpha_{T}$ & -4.5 & -4 & -3.9 & \\ \hline
    $\ell_{\textnormal{knee},P}$ & 300 & 490 & 500 & \\ \hline
    $\alpha_{P}$ & -2.2 & -2 & -2.5 & \\ \hline\hline
    \multicolumn{5}{l}{SO Standard:} \\
    \hline
    $f_\mathrm{sky}$ & \multicolumn{4}{c}{0.4} \\ \hline
    $\sigma$ [arcmin] & 2.2 & 1.4 & 1.0 & 0.9 \\ \hline
    $\Delta_{T}$ [$\mu$K-arcmin] & 5.8 & 6.3 & 15 & 37 \\ \hline
    $\ell_{\textnormal{knee},T}$ & \multicolumn{4}{c}{1000} \\ \hline
    $\alpha_{T}$ & \multicolumn{4}{c}{-3.5} \\ \hline
    $N_{\textnormal{red},T}$ [$\mu$K$^2$s] & 230 & 1500 & 17000 & 31000 \\ \hline
    $\ell_{\textnormal{knee},P}$ & \multicolumn{4}{c}{700} \\ \hline
    $\alpha_{P}$ & \multicolumn{4}{c}{-1.4} \\ \hline\hline
    \multicolumn{5}{l}{SO Extended:} \\
    \hline
    $f_\mathrm{sky}$ & \multicolumn{4}{c}{0.4} \\ \hline
    $\sigma$ [arcmin] & 2.2 & 1.4 & 1.0 & 0.9 \\ \hline
    $\Delta_{T}$ [$\mu$K-arcmin] & 3.8 & 4.1 & 10 & 25 \\ \hline
    $\ell_{\textnormal{knee},T}$ & 2100 & 3000 & 6740 & 6792 \\ \hline
    $\alpha_{T}$ & \multicolumn{4}{c}{-3.5} \\ \hline
    $\ell_{\textnormal{knee},P}$ & \multicolumn{4}{c}{700} \\ \hline
    $\alpha_{P}$ & \multicolumn{4}{c}{-1.4} \\ \hline\hline
    \end{tabular}
    \centering \caption{Experimental specifications used in this work. Here, $f_\mathrm{sky}$ is sky fraction, $\sigma$ is beam FWHM, $\Delta_T$ is temperature white noise level, and $\ell_{\textnormal{knee}}$, $\alpha$ are the $1/f$ parameters. Polarization noise levels are assumed to be $\sqrt{2}$ times $\Delta_T$.}
    \label{tab:experimental_specs}
\end{table}

In Table~\ref{tab:experimental_specs} we summarize the experimental specifications used in this work, as described also in Section~\ref{sec:noise_levels}. The noise spectrum is modeled as
\begin{equation}\label{eq:noise_model}
    N_\ell = N_\textnormal{red} \left(\frac{\ell}{\ell_\textnormal{knee}}\right)^\alpha + N_\textnormal{white},
\end{equation}
where we assume $N_\textnormal{red} = N_\textnormal{white}$ unless otherwise stated.

\section{tSZ-CIB Interactions} \label{sec:tszxcib}

\begin{figure}[t]
	\centering
	\includegraphics[width=.6\textwidth]{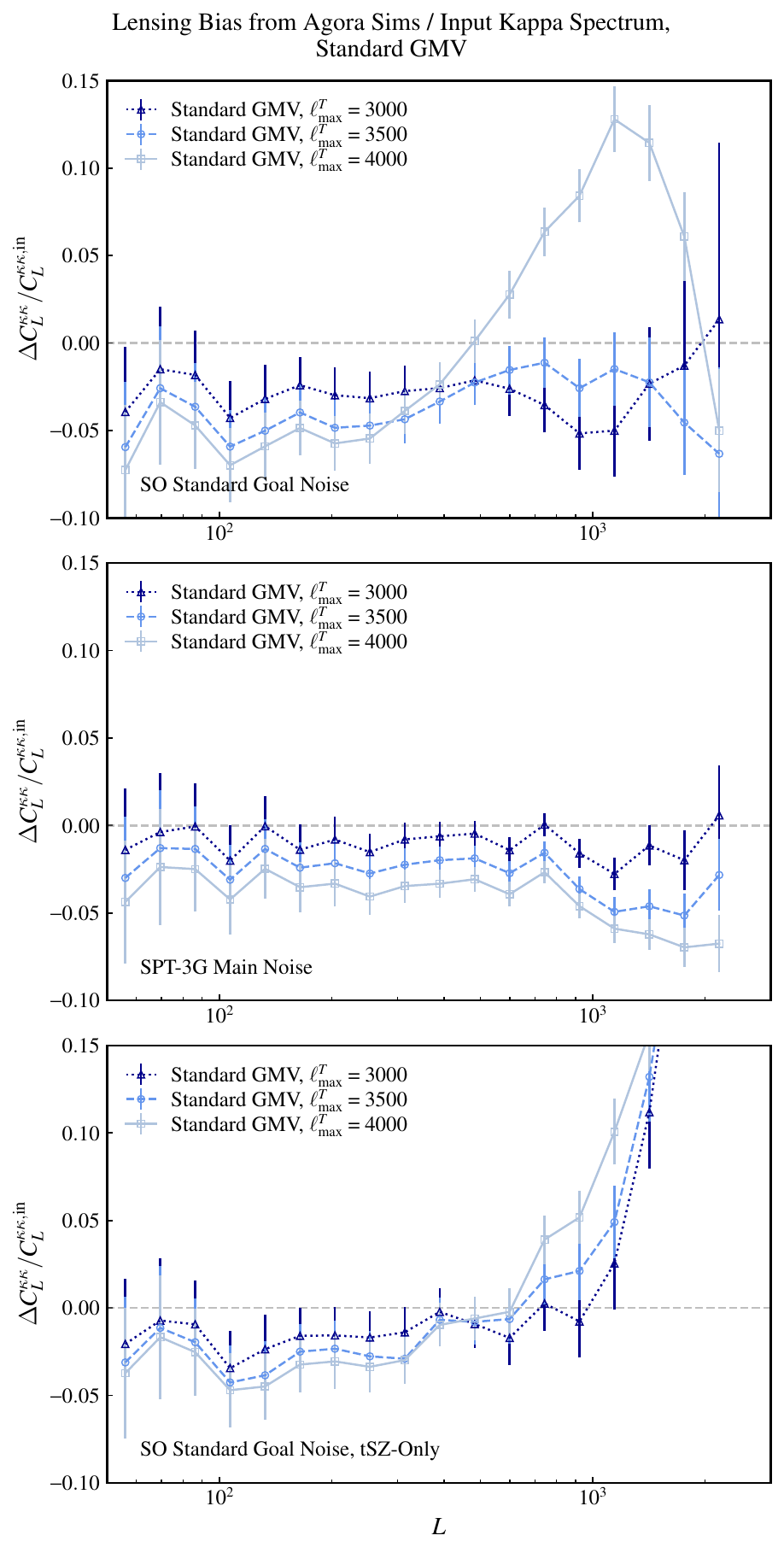}
	\centering \caption{Lensing bias for standard reconstruction for varying $\ell_{\mathrm{max}}^T$.
	The error bars are measurement error for full sky.
	Top panel: SO Standard depth goal noise case. We observe the $\ell_{\mathrm{max}}^T = 4000$ has an anomalous bump, peaking at around $L = 1000$.
	Middle panel: SPT-3G Main noise case. We observe the $\ell_{\mathrm{max}}^T = 4000$ bump is not present. SPT-3G Main noise levels are the lowest out of the experiments considered in this work.
	Bottom panel: SO Standard depth goal noise case, where only tSZ is present as the non-Gaussian foreground in the \agora\ simulations. The anomalous $\ell_{\mathrm{max}}^T = 4000$ behavior is now gone. We see a similar result with CIB-only instead of tSZ-only.}
	\label{fig:bias_standard_different_lmaxT_stacked}
\end{figure}

We find that there is a ``bump" in the lensing bias, in which the bias becomes less negative (or more positive) around $L \sim 1000$ before trending downward again, for the standard (i.e., no foreground treatment) case at $\ell_{\mathrm{max}}^T = 4000$. This feature appears in every noise scenario except for SPT-3G Main (which has the lowest noise levels out of all experiments considered in this work). The most obvious example of this is shown in the top subplot of Figure~\ref{fig:bias_standard_different_lmaxT_stacked} for SO Standard depth noise, which has the highest experimental noise. In contrast, with SPT-3G Main noise levels, we see no bump at all; see the middle plot in Figure~\ref{fig:bias_standard_different_lmaxT_stacked}.

To investigate this anomaly, we performed lensing reconstructions using \agora\ simulations with only a single non-Gaussian foreground component included at a time. With SO Standard noise levels, the tSZ-only case yields the lensing bias shown in the bottom subplot of Figure~\ref{fig:bias_standard_different_lmaxT_stacked}, while the CIB-only case produces a similar result. In both of these cases, the anomalous feature at $\ell_{\mathrm{max}}^T = 4000$ is absent. However, when both tSZ and CIB are included simultaneously, the feature reappears. This points to tSZ–CIB correlations as the likely source of this behavior. To confirm this, we performed an additional test in which both tSZ and CIB are included but the CIB map is rotated, so that the tSZ-CIB correlation is broken. In this case, the feature disappears, further supporting the interpretation that mixed tSZ–CIB correlations are the source of the anomaly.

We have also explored the use of the CosmoBLENDER code~\cite{anton_halo_model} to model foreground-induced lensing biases within the halo model framework, applying halo mass cuts chosen to roughly match the cluster masking used in our analysis. In these tests, we did not reproduce the anomalous lensing bias seen in our simulations when both non-Gaussian tSZ and CIB foregrounds are present. However, with CosmoBLENDER we model the foregrounds only at leading order in the halo model, while the \agora\ simulations in our analysis naturally include higher order correlations. Thus, the absence of the effect in the halo model prediction does not rule out the possibility that the bias induced by tSZ-CIB correlations observed in \agora\ represents a genuine physical effect.

\acknowledgments

We would like to thank Ant\'on Baleato Lizancos, Srini Raghunathan, Federico Bianchini and Dominic Beck for useful discussions during the course of this work. We also thank Yunyang Li for helpful comments on an earlier version of this paper.

Much of the computing for this project was performed on the Sherlock cluster. We would like to thank Stanford University and the Stanford Research Computing Center for providing computational resources and support that contributed to these research results. We also acknowledge the computing resources provided
on Crossover, a high-performance computing cluster operated by the
Laboratory Computing Resource Center at Argonne National Laboratory.
WLKW acknowledges support from an Early Career Research Award of the Department of Energy and a Laboratory Directed Research and Development program as part of the Panofsky Fellowship program at the SLAC National Accelerator Laboratory.

\bibliographystyle{JHEP}
\bibliography{biblio.bib}

\end{document}